 \documentclass[11pt]{article}

\usepackage{amsmath,amssymb}

\usepackage[all]{xy}

\newtheorem{definition}{{\bf Definition}}

\newtheorem{theorem}{Theorem}
\newtheorem{proposition}{{\bf Proposition}}

 1   
 1  

\def\qed{\quad\rule{8pt}{8pt}}
\newenvironment{proof}{\noindent{({\it Proof\/}):\quad~}}{\qed\medskip}
\def\QED{\hskip0.1em\hfill\null\ \null\nobreak\hfill
\kern3pt\lower1.8pt\vbox{\hrule\hbox {\vrule\kern1pt\vbox{\kern1.7pt
\hbox{$\scriptstyle QED$}\kern0.2pt}\kern1pt\vrule}\hrule}}
 \def\hook{\, \hbox to 15pt{\vbox{\vskip 6pt\hrule width 8pt height 1pt}
         \kern -5pt\vrule height 8pt width 1pt\hfil}}

\def\Real{\mathbb{R}}
\def\r{\ensuremath{\mathbb{R}}}
\def\rk{{\mathbb R}^{k}}
\def\rkq{\rk \times Q}

\def\Lagd{\mathbb{L}}
\def\Lag{\cal L}
\def\beq{\begin{equation}}
\def\eeq{\end{equation}}
\def\bea{\begin{eqnarray}}
\def\eea{\end{eqnarray}}
\def\beann{\begin{eqnarray*}}
\def\eeann{\end{eqnarray*}}
\def\ben{\begin{enumerate}}
\def\een{\end{enumerate}}
\def\bit{\begin{itemize}}
\def\eit{\end{itemize}}
\newcommand{\ds}{\displaystyle}
\newcommand{\inn}{\langle\,,\,\rangle}

\def\d{{\rm d}}
\def\vf{\mathfrak{X}}
\def\df{{\mit\Omega}}
\def\Tan{{\rm T}}
\def\Lie{\mathop{\rm L}\nolimits}
\def\inn{\mathop{i}\nolimits}
\def\Cinfty{{\rm C}^\infty}
\def\derpar#1#2{\frac{\partial{#1}}{\partial{#2}}}

\def\tabaddress#1{{\small\it\begin{tabular}[t]{c}#1
\\[1.2ex]\end{tabular}}}
\def\qed{\ifvmode\removelastskip\fi
{\unskip\nobreak\hfil\penalty50\hbox{}\nobreak\hfil \hbox{\vrule
height1.2ex width1.2ex}\parfillskip=0pt \finalhyphendemerits=0
\par\smallskip}}

\title{ON THE $k$-SYMPLECTIC, $k$-COSYMPLECTIC AND
MULTISYMPLECTIC FORMALISMS OF CLASSICAL FIELD THEORIES}
\author{\sc Narciso Rom\'an-Roy\thanks{{\bf e}-{\it mail}:
  nrr@ma4.upc.edu}
  \\
  \tabaddress{Departamento de Matem\'atica Aplicada IV.
  Edificio C-3, Campus Norte UPC\\
  C/ Jordi Girona 1. 08034 Barcelona. Spain}
  \\
{\sc  \'Angel M. Rey\thanks{{\bf e}-{\it mail}: angelmrey@edu.xunta.es},
 Modesto Salgado\thanks{{\bf e}-{\it mail}: modesto@zmat.usc.es},
     Silvia Vilari\~no\thanks{{\bf e}-{\it mail}:
svfernan@usc.es}} \\
 \tabaddress{Departamento de Xeometr{\'\i}a e Topolox{\'\i}a\\
 Facultade de Matem\'{a}ticas,
    Universidade de Santiago de Compostela,\\
    15706-Santiago de Compostela, Spain}}

\begin{document}

\maketitle

\pagestyle{myheadings}
\parskip=7pt

\thispagestyle{empty}

\begin{abstract}
The objective of this work is twofold:
First, we analyze the relation between the
$k$-cosymplectic and the $k$-symplectic Hamiltonian and Lagrangian
formalisms in classical field theories.
In particular, we prove the equivalence between
$k$-symplectic field theories and
the so-called autonomous $k$-cosymplectic field theories,
extending in this way the description of the symplectic formalism of autonomous systems as
a particular case of the cosymplectic formalism in non-autonomous mechanics.
Furthermore, we clarify some aspects of the
geometric character of the solutions to the
Hamilton-de Donder-Weyl and the Euler-Lagrange equations
in these formalisms.
Second, we study the equivalence between
$k$-cosymplectic and a particular kind of multisymplectic Hamiltonian and Lagrangian
field theories (those where the configuration bundle of the theory is trivial).
\end{abstract}

\bigskip \bigskip\bigskip


\vbox{\raggedleft AMS s.\,c.\,(2000): 70S05, 53D05, 53D10}\null

\markright{\sc N. Rom\'an-Roy {\it et al\/},
    \sl $k$-symplectic, $k$-cosymplectic and  multisymplectic formalisms\ldots}

  \clearpage

\tableofcontents

\section{Introduction}

The {\sl $k$-symplectic} and {\sl $k$-cosymplectic formalisms}
are the simplest geometric frameworks for describing classical field theories.
The $k$-symplectic formalism \cite{gun,fam}
(also called {\sl polysymplectic formalism}\/) is the generalization
to field theories of the standard symplectic formalism in
autonomous mechanics, and is used to give a geometric description of certain kinds
of field theories: in a local description, those whose Lagrangian and Hamiltonian functions
do not depend on the coordinates in the basis (in many of them,
the space-time coordinates). The foundations of the
$k$-symplectic formalism are  the $k$-symplectic manifolds
intoduced in \cite{aw,aw2,aw3}.
The $k$-cosymplectic formalism is the generalization to field
theories of the standard cosymplectic formalism for non-autonomous mechanics,
 \cite{mod1,mod2}, and it describes field
theories involving the coordinates in the basis on the Lagrangian
 and on the Hamiltonian. The foundations of the $k$-cosymplectic formalism
are the $k$-cosymplectic manifolds introduced in \cite{mod1,mod2}.
 One of the advantages of these formalisms
 is that only the tangent and cotangent bundle of a
manifold are required for their development.
(A brief review of $k$-symplectic and $k$-cosymplectic geometry is given in Section \ref{ksc}).
Other different polysymplectic formalisms
for describing field theories have been proposed in
\cite{Sarda2,Sarda1,Kana,McN,No2,No5,Sd-95}.

In these formalisms, the field equations (Hamilton-de Donder-Weyl and Euler-Lagrange equations)
can be written in a geometrical way using integrable $k$-vector fields.
However, although integral sections of integrable $k$-vector fields
(i.e., integrable distributions) that are solutions to
the geometrical field equations are proved to be solutions to the
Hamilton-de Donder-Weyl or the Euler-Lagrange equations,
the converse is not always true.
This also occurs when other geometric descriptions of classical field theories
in terms of multivector fields are considered
 (see \cite{EMR-98,EMR-99b,PR-2002} for details in the case of
multisymplectic field theories).
Here we prove that, in the $k$-cosymplectic formalism, every solution to
the Hamilton-de Donder-Weyl equations is, in fact, an integral section of an
integrable $k$-vector field that is a solution to the geometrical field equations
in the Hamiltonian formalism.
Nevertheless,  in the $k$-symplectic Hamiltonian formalism, this is no longer true,
unless some additional conditions on the solutions to the
Hamilton-de Donder-Weyl are required.
All these features are discussed in Sections \ref{kshs}, \ref{kchs1}, \ref{akcs}, \ref{kslf},
and  \ref{kcolag}.

After reviewing the $k$-cosymplectic Hamiltonian formalism in Section  \ref{kchs1},
Section \ref{akcs} contains other relevant results of this work.
In particular, the relation between the
$k$-cosymplectic and the $k$-symplectic Hamiltonian formalism
is studied here, proving the equivalence between
$k$-symplectic Hamiltonian systems and a class of  $k$-cosymplectic Hamiltonian systems:
the so-called {\sl autonomous $k$-cosymplectic Hamiltonian systems}.
This generalizes the situation in classical mechanics,
where the symplectic formalism for describing autonomous Hamiltonian systems
can be recovered as a particular case of the cosymplectic Hamiltonian formalism
when systems described by time-independent Hamiltonian functions are considered.

A more general geometric framework for describing classical field theories
is the {\sl multisymplectic formalism} \cite{CCI-91,GIMMSY-mm,MS-99},
first introduced in \cite{Ki-73,KS-75,KT-79},
 which is based on the use of multisymplectic manifolds.
In particular, jet bundles are the appropriate domain for
stating the Lagrangian formalism \cite{Sa-89},
and different kinds of multimomentum bundles are used for
developing the Hamiltonian description
\cite{EMR-00,HK-04,LMM-96}.
(A brief review of multisymplectic Hamiltonian and Lagrangian
field theories is given in Sections \ref{mmb}, \ref{hsjpistar}, and \ref{mls}).

Multisymplectic models allow us to describe a higher variety of field theories
than the $k$-cosymplectic or $k$-symplectic models,
since for the latter the configuration bundle of the theory
must be a trivial bundle; however, this restriction does not oocur for the former.
Another goal of this paper is to show the equivalence between
 the multisymplectic and $k$-cosymplectic descriptions,
 when theories with trivial configuration bundles are considered,
for both the Hamiltonian and Lagrangian formalisms.
In this way we complete the results obtained in \cite{LMNRS-2002},
where an initial analysis about the relation between multisymplectic, $k$-cosymplectic and
$k$-symplectic structures was carried out.
This study is explained in Sections \ref{rkcf}, and \ref{rmk}.

All manifolds are real, paracompact, connected and $C^\infty$. All
 maps are $C^\infty$. Sum over crossed repeated indices is understood.

\section{$k$-symplectic and $k$-cosymplectic Hamiltonian formalisms}

\subsection{k-vector fields and integral sections}
\protect\label{kvfis}

(See  \cite{mod1} and \cite{RRS} for details).
  If $M$ is a differentiable manifold, let $T^{1}_{k}M=TM\oplus \stackrel{k}{\dots} \oplus TM$ be
the Whitney sum of $k$ copies of $TM$, and $\tau_M^1 \colon
T^{1}_{k}M \longrightarrow M$ its  canonical projection.
$T^{1}_{k}M$ is usually called the {\sl $k$-tangent bundle} or {\sl tangent bundle of
$k^1$-velocities of $M$}.

\begin{definition}
\label{kvector} A {\rm $k$-vector field} on $M$ is a section
 ${\bf X} \colon M \longrightarrow T^1_kM$
of the projection $\tau_M^1$.
\end{definition}

Giving a $k$-vector field ${\bf X}$ is equivalent to
giving a family of $k$ vector fields $X_{1}, \dots, X_{k}$ on $M$
obtained by projecting ${\bf X}$ onto every factor; that is,
$X_A=\tau_A\circ{\bf X}$, where $\tau_A\colon T^1_kM \rightarrow
TM$ is the canonical projection onto the $A^{th}$-copy $TM$ of
$T^1_kM$. For this reason we will denote a $k$-vector field by
${\bf X}=(X_1, \ldots, X_k)$.

\begin{definition} \label{integsect}
An {\rm integral section} of the $k$-vector field ${\bf
X}=(X_{1},\dots, X_{k})$
 passing through a point $x\in M$  is a map
  $\phi\colon U_0\subset \r^k \rightarrow M$, defined on some neighborhood
 $U_0$ of $0\in \rk$,  such that
$$
\phi(0)=x, \, \,
\phi_{*}(t)\left(\frac{\displaystyle\partial}{\partial
t^A}\Big\vert_t\right) = X_{A}(\phi (t))
 \quad , \quad \mbox{\rm for every $t\in U_0$, $1\leq A \leq k$} \, .
$$
A $k$-vector field is said to be {\rm integrable} if there is an integral
section passing through every point of $M$.
\end{definition}

{\bf Remark}: $k$-vector fields in a manifold ${\cal M}$ can also be
defined more generally as sections of the bundle $\Lambda^k({\rm T}
{\cal M})\to {\cal M}$ (i.e., the contravariant skew-symmetric
tensors of order $k$ in ${\cal M}$). The $k$-vector fields defined
in Definition \ref{kvector} are a particular class: the so-called
{\sl decomposable} or {\sl homogeneous $k$-vector fields}, which can
be associated with distributions on ${\cal M}$. We remark that a
$k$-vector field ${\bf X}=(X_1,\ldots , X_k)$ is integrable if, and
only if, $\{ X_1,\ldots , X_k\}$ define an involutive distribution
on ${\cal M}$. (See \cite{EMR-98} for a detailed exposition on these
topics).

\subsection{$k$-symplectic and $k$-cosymplectic manifolds}
\protect\label{ksc}

(See  \cite{mod1} and \cite{RRS} for details).

\begin{definition}\label{defaw} {\rm (Awane \cite{aw})}
 A {\rm $k$-symplectic structure} on  a manifold $M$ of dimension $N=n+kn$
is a  family $(\omega^A,V;1\leq A\leq k)$, where each $\omega^A$
is a closed $2$-form and $V$ is an integrable $nk$-dimensional
distribution on $M$ such that
 $$
(i) \quad \omega^A\vert_{ V\times V}=0,\qquad
 (ii) \quad \cap_{A=1}^{k} \ker\omega^A=\{0\} \ .
$$
Then $(M,\omega^A,V)$ is called a {\rm $k$-symplectic manifold}.
\end{definition}

\begin{theorem}{\rm (Awane \cite{aw})}
 Let $(\omega^A,V;1\leq A\leq k)$ be a $k$-symplectic
structure on $M$. For every point of $M$  there exists a local
chart of coordinates $(q^i , p^A_i),\, 1\leq i \leq n,\, 1\leq A
\leq k$, such that
$$
\omega^A=  dq^i\wedge dp^A_{i} \quad , \quad
V=\left\langle\frac{\partial}{\partial p^1_i}, \dots,
\frac{\partial}{\partial p^k_i}\right\rangle_{i=1,\ldots , n}\quad
; \quad  1\leq A \leq k \quad .
$$
\end{theorem}

The canonical model for this geometrical structure is
$((T^1_k)^*Q,\omega^A,V)$, where $Q$ is a $n$-dimensional
differentiable manifold
 and $(T^1_k)^*Q= T^*Q \oplus \stackrel{k}{\dots} \oplus
T^*Q$ is the Whitney sum of $k$ copies of the cotangent bundle
$T^*Q$, which is usually called the {\sl $k$-cotangent bundle} or {\sl bundle of
$k^1$-covelocities} of $Q$. We use the following notation for the
canonical projections:
$$
\pi^A\colon (T^1_k)^*Q \rightarrow T^*Q\quad , \quad \pi^1_Q\colon
(T^1_k)^*Q \to Q\quad ;\quad
 \quad (1\leq A\leq k) \ ,
$$
(here $\pi^A$ is the canonical projection onto the $A^{th}$-copy
$T^*Q$ of $(T^1_k)^*Q$). So, if $q\in Q$ and $(\alpha^1_q,
\ldots ,\alpha^k_q)\in (T^1_k)^*Q$, we have
$$
\pi^A(\alpha^1_q, \ldots ,\alpha^k_q)=\alpha^A_q \quad ,  \quad
 \pi^1_Q(\alpha^1_q, \ldots ,\alpha^k_q)=q \quad (1\leq A\leq k) \ .
$$
If $(q^i)$, $1\leq i\leq n$, are local coordinates on $U \subseteq Q$, the
induced local coordinates  $(q^i ,p^A_i)$ on
$(\pi^1_Q)^{-1}(U)=(T^1_k)^*U$ are given by
$$
q^i(\alpha^1_q, \ldots ,\alpha^k_q)= q^i(q)\quad , \quad
p^A_i(\alpha^1_q, \ldots ,\alpha^k_q)=\alpha^A_
q\left(\frac{\partial}{\partial q^i}\Big\vert_q\right) \ .
$$

The canonical $k$-symplectic structure in $(T^1_k)^*Q$ is constructed as follows:
we define the differential forms
 \begin{equation}
 \theta^A=(\pi^A)^*\theta\quad , \quad
 \omega^A=(\pi^A)^*\omega \quad ; \quad 1\leq A\leq k \ ,
\label{symforms}
 \end{equation}
 where $\theta$ is the Liouville $1$-form on $T^*Q$ and
$\omega=-\d\theta$ is the canonical symplectic form on $T^*Q$.
Obviously $\omega^A = -d\theta^A$. In local coordinates we have
\begin{equation}
\label{locexp0} \theta^A = p^A_i\d q^i \quad  ,  \quad \omega^A =
\d q^i\wedge\d p^A_i\quad ;\quad 1\leq A\leq k \ .
\end{equation}
The canonical $k$-symplectic manifold is $((T^1_k)^*Q,\omega^A, V)$
where $V=\ker \, (\pi^1_Q)_*$.

\begin{definition}
\label{deest}
 Let $M$ be a differentiable manifold  of dimension $k(n+1)+n$.
A {\rm $k$--cosymplectic structure} is a family
$(\eta^A,\Omega^A,{\cal V})$ ($1\leq A\leq k$), where
$\eta^A\in\df^1(M)$, $\Omega^A\in\df^2(M)$, and ${\cal V}$ is an
$nk$-dimensional distribution on $M$, such that
\begin{enumerate}
\item
 $\eta^1\wedge\dots\wedge\eta^k\neq 0$, \quad
$\eta^A\vert_{\cal V}=0,\quad\Omega^A\vert_{{\cal V}\times {\cal
V}}=0$.
 \item
 $({\cap_{A=1}^{k}} \ker\eta^A) \cap ( {\cap_{A=1}^{k}}\ker\Omega^A)=\{0\}$,
\quad $dim({\cap_{A=1}^{k}}\ker\Omega^A)=k$.
\item
The forms $\eta^A$ and $\Omega^A$ are closed, and
${\cal V}$ is integrable.
\end{enumerate}
Then, $(M,\eta^A,\Omega^A,{\cal V})$ is said to be a {\rm $k$--cosymplectic manifold}.
\end{definition}

For every $k$-cosymplectic structure  $(\eta^A ,\Omega^A,{\cal
V})$ on $M$, there exists a family of $k$ vector fields
$\{R_A\}_{\, 1\leq A\leq k}$, which are called {\sl Reeb vector fields},
 characterized by the following conditions
$$
\inn(R_A)\eta^B=\delta^B_A \quad ,\quad \inn(R_A)\Omega^B=0\quad ;
\quad 1\leq A,B \leq k \ .
$$

\begin{theorem}  {\rm (Darboux Theorem)}:
If $M$ is a $k$--cosymplectic manifold, then for every point of
$M$ there exists a local chart of coordinates $(t^A,q^i,p^A_i)$,
$1\leq A\leq k$, $1\leq i \leq n$, such that
$$
\eta^A=\d t^A,\quad \Omega^A=\d q^i\wedge\d p^A_i, \quad {\cal
V}=\left\langle\frac{\partial} {\partial p^1_i}, \dots,
 \frac{\partial}{\partial p^k_i}\right\rangle_{i=1,\ldots , n}.
$$
\end{theorem}

The canonical model for these geometrical structures is
$(\rk\times (T^1_k)^*Q,\eta^A,\Omega^A,{\cal V})$.
If $(t^A)$ are
coordinates in $\rk$, and $(q^i)$ are local coordinates on $U
\subset Q$, then the induced local coordinates
$(t^A,q^i,p^A_i)$ on $\rk \times (T^1_k)^*U$ are given by
$$
 t^A(t,\alpha^1_q,\ldots ,\alpha^k_q) = t^A\quad , \quad
 q^i(t,\alpha^1_q, \ldots ,\alpha^k_q) = q^i(q)\quad , \quad
  p_i^A(t,\alpha^1_q, \ldots ,\alpha^k_q) =
 \alpha^A_q\left(\ds\frac{\partial}{\partial q^i}\Big\vert_q
 \right)\, .
$$

Considering the canonical projections (submersions), we have the commutative diagram:

\begin{equation}
\xymatrix{
\rk\times Q & & \rk\times (T^1_k)^*Q
\ar[ll]_{\txt{\small{$\bar\pi_0$}}}
\ar[rr]^(.6){\txt{\small{$\bar\pi_k$}}}
\ar[dd]^(.5){\txt{\small{$\bar\pi^A$}}}
\ar[ddddll]_{\txt{\small{$\bar\pi_2$}}}
\ar@/^2pc/[rrrr]^(.7){\txt{\small{$\bar\pi_k^A$}}}
\ar@/_2.5pc/[dddd]_(.7){\txt{\small{$\bar\pi^A_2$}}}
\ar@/^16pc/[ddddrr]_(.5){\txt{\small{$\bar\pi^1_Q$}}} &  & \rk
\ar[rr]^(.4){\txt{\small{$t^A$}}} & & \r
\\ \\
& & \r\times T^*Q
\ar[uurrrr]^(.5){\txt{\small{$\pi^A_k$}}}
\ar[rr]^(.5){\txt{\small{$\pi_0$}}}
\ar[dd]^{\txt{\small{$\pi_2$}}} & &\r\times Q
\ar[uurr]_{\txt{\small{$\rho_1$}}}
\ar[dd]^{\txt{\small{$\rho_2$}}} & &
\\ \\
 (T^1_k)^*Q
\ar[rr]^{\txt{\small{$\pi^A$}}}
\ar@/_1pc/[rrrr]_{\txt{\small{$\pi^1_Q$}}} & & T^*Q
\ar[rr]^{\txt{\small{$\pi_Q$}}} & & Q }
\label{megadiag}
\end{equation}
 In particular, if  $t=(t^1,\ldots,t^k)\in\rk $, $q\in Q$ and
$(t,\alpha^1_q, \ldots ,\alpha^k_q)\in\rk\times(T^1_k)^*Q$,
 we have
$$
\begin{array}{ccccccc}
\bar\pi_2(t,\alpha^1_q, \ldots ,\alpha^k_q)&=&(\alpha^1_q,
\ldots ,\alpha^k_q) & , & \bar\pi^A_2(t,\alpha^1_q, \ldots
,\alpha^k_q)&=&(\alpha^A_q)
\\
\bar\pi^1_Q(t,\alpha^1_q, \ldots ,\alpha^k_q)&=&q  & , &
\bar\pi^A_k(t,\alpha^1_q, \ldots ,\alpha^k_q)&=&t^A
\\
\bar\pi^k(t,\alpha^1_q, \ldots ,\alpha^k_q)&=& t & , &
\bar\pi^A(t,\alpha^1_q, \ldots ,\alpha^k_q)&=&(t^A,\alpha^A_q)
\end{array}
$$

The canonical $k$-cosymplectic structure in $\rk\times (T^1_k)^*Q$ is constructed as
follows:  we define the differential forms
 \begin{equation}
 \eta^A=(\bar\pi^A_k)^*\d t^A\quad , \quad
 \Theta^A=(\bar\pi^A_2)^*\theta\quad , \quad
 \Omega^A= (\bar\pi^A_2)^*\omega\quad ; \quad 1\leq A \leq k \ .
\label{cosymforms}
 \end{equation}
 Obviously $\Omega^A =-d\Theta^A$. In local coordinates we have
 \begin{equation}
 \label{lockcos}
   \eta^A=\d t^A \quad , \quad
\Theta^A = p^A_i\d q^i \quad  ,  \quad
 \Omega^A = \d q^i\wedge\d p^A_i\quad ;\quad  1\leq A \leq k
 \end{equation}

The canonical $k$-cosymplectic manifold is $(\rk\times
(T^1_k)^*Q,\eta^A,\Omega^A,{\cal V})$ where ${\cal
V}=\ker\,(\bar\pi_0)_*$, and locally $\ds{\cal
V}=\left\langle\derpar{}{ p^A_i}\right\rangle_{1\leq A\leq k,\ 1\leq
i\leq n}$. Moreover, the Reeb vector fields are $\ds R_A=\derpar{}{
t^A}$, $1\leq A\leq k$, which are defined intrinsically in
$\rk\times (T^1_k)^*Q$ and span locally the vertical distribution
with respect to the projection $\bar\pi_2$; i.e., the distribution
generated by $\ker\,(\bar\pi_2)_*$.

Finally, taking into account (\ref{symforms}), (\ref{cosymforms}),
and the commutativity of the diagram (\ref{megadiag}), we have
that
 \begin{equation}
 \Theta^A=\bar\pi_2^*\theta^A \quad , \quad
\Omega^A=\bar\pi_2^*\omega^A \quad ; \quad 1\leq A\leq k\, .
\label{symcosym}
 \end{equation}
 Furthermore, the vector fields
spanning the distributions ${\cal V}$ on $\rk\times(T^1_k)^*Q$,
and $V$ on $(T^1_k)^*Q$ are also $\bar\pi_2$-related.

\subsection{$k$-symplectic Hamiltonian systems}
\protect\label{kshs}

Consider the  $k$-symplectic manifold
$((T^1_k)^*Q,\omega^A,V)$, and
 let $H\in\Cinfty((T^1_k)^*Q)$ be a Hamiltonian function.
$((T^1_k)^*Q,H)$ is called a {\sl $k$-symplectic Hamiltonian
system}. The {\sl Hamilton-de Donder-Weyl equations} (HDW-equations
for short) for this system are the set of partial diferential
equations:
\begin{equation}
\label{he20}
 \frac{\partial H}{\partial q^i}=-\ds\sum_{A=1}^k\frac{\partial\psi^A_i} {\partial t^A}
\quad , \quad \frac{\partial H} {\partial
p^A_i}=\frac{\partial\psi^i}{\partial t^A},\quad ;\quad \quad 1\leq
i\leq n, \ 1\leq\ A \leq k\, ,
\end{equation}
where $\psi\colon\Real^k\to (T^1_k)^*Q$,
$\psi(t)=(\psi^i(t),\psi^A_i(t))$, is a solution.

We denote by $\vf^k_H((T^1_k)^*Q)$ the set of $k$-vector fields
${\bf X}=(X_1,\dots,X_k)$ on $(T^1_k)^*Q$ which are solutions to the
equations
\begin{equation}
\ds\sum_{A=1}^k\inn(X_A)\omega^A=\d H\;.
 \label{generic0}
\end{equation}

 In a local system of canonical coordinates, each $X_A$ is locally given by
 \begin{equation}
  \label{coorxa}
   X_A =(X_A)^i\frac{\partial}{\partial q^i}+(X_A)_i^B\frac{\partial}{\partial p_i^B} 
\quad , \quad \ 1\leq\ A \leq k\ ,
\end{equation}
then, using (\ref{locexp0}), we obtain that the equation
(\ref{generic0}) is equivalent to the equations
 \begin{equation}
  \label{11}
 \frac{\partial H}{\partial q^i}=\, -\ds\sum_{A=1}^k\,(X_A)^A_i
 \quad  , \quad \frac{\partial H}{\partial p^A_i}=(X_A)^i  \quad , \quad 1\leq i\leq n\ .
\end{equation}

The existence of $k$-vector fields that are solutions to (\ref{generic0})
 is assured, and in a local system of coordinates they depend on
$n(k^2-1)$ arbitrary functions.
Nevertheless, they are not necessarily integrable,
and hence the integrability conditions imply that
the number of arbitrary functions will
in general be less than $n(k^2-1)$.

\begin{proposition}
 Let ${\bf X}=(X_1,\dots,X_k)$ be an integrable $k$-vector field in $(T^1_k)^*Q$
and $\psi\colon\Real^k\to (T^1_k)^*Q$ an integral section of ${\bf
X}$. Then $\psi(t)=(\psi^i(t),\psi^A_i(t))$ is a solution to the
HDW-equations (\ref{he20}) if, and only if, ${\bf
X}\in\vf^k_H((T^1_k)^*Q)$.
 \label{prop1}
\end{proposition}
\proof
If $\psi(t)=(\psi^i(t),\psi^A_i(t))$ is an integral section of ${\bf X}$, then
 \begin{equation}
  \label{intsec0}
\frac{\partial\psi^i}{\partial t^B}=(X_B)^i \quad , \quad
\frac{\partial\psi^A_i}{\partial t^B}=(X_B)^A_i \,  .
\end{equation}
and therefore (\ref{11}) are the
 HDW-equations (\ref{he20}).
\qed

{\bf Remark}: It is important to point out that the equations
(\ref{he20}) and (\ref{generic0}) are not equivalent, because there
is no way to prove that every solution to the HDW-equations
(\ref{he20}) is an integral section of some integrable $k$-vector
field of $\vf^k_H((T^1_k)^*Q)$, unless some additional conditions
are required. In particular, we could assume the following condition
(which holds for a large class of mathematical applications and
physical field theories):

\begin{definition}
\label{hyp} A map $\psi\colon\rk\to (T^1_k)^*Q$, solution to the
equations (\ref{he20}),
 is said to be an {\rm admissible solution} to the HDW-equations
for a $k$-symplectic Hamiltonian system $((T^1_k)^*Q,H)$, if ${\rm
Im}\,\psi$ is a closed embedded submanifold of $(T^1_k)^*Q$.

We say that $((T^1_k)^*Q,H)$  is an {\rm admissible $k$-symplectic
Hamiltonian system} if all the solutions to its HDW-equations are
admissible.
\end{definition}

\begin{proposition}
Every admissible solution to the HDW-equations (\ref{he20}) is an
integral section of an integrable $k$-vector field ${\bf
X}\in\vf^k_H((T^1_k)^*Q)$.
 \label{prop3}
\end{proposition}
\proof
 Let $\psi\colon\rk\to(T^1_k)^*Q$ be an admissible solution
to the HDW-equations (\ref{he20}). By hypothesis, ${\rm Im}\,\psi$
is a $k$-dimensional closed submanifold of $(T^1_k)^*Q$. As $\psi$
is an embedding, we can define a $k$-vector field ${\bf
X}\vert_{{\rm Im}\,\psi}$ (at support on ${\rm Im}\,\psi$), and
tangent to ${\rm Im}\,\psi$, by
$$
 X_A(\psi(t))=(\psi)_*(t)\left(\frac{\partial}{\partial
t^A}\Big\vert_t\right)
$$
which is a solution to (\ref{generic0}) on the points of ${\rm
Im}\,\psi$, since (\ref{11}) holds on these points as a consequence
of (\ref{he20}) and (\ref{intsec0}).
 Furthermore, by hypothesis, ${\rm Im}\,\psi$ is a
closed submanifold of $(T^1_k)^*Q$; therefore we can extend  this
$k$-vector field ${\bf X}\vert_{{\rm Im}\,\psi}$ to an integrable
$k$-vector field  ${\bf X}\in\vf^k_H((T^1_k)^*Q)$ in such a way that
this extension is a solution to the equations (\ref{generic0})
(remember that these equations have solutions
everywhere on $(T^1_k)^*Q$), and which obviously
has $\psi$ as an integral section. This extension is made at least
locally, and then the global $k$-vector field is constructed using
partitions of unity.
 \qed

In this way, for admissible $k$-symplectic Hamiltonian systems,
 the field equations (\ref{generic0}) are a geometric
version of the HDW-equations (\ref{he20}).

\subsection{$k$-cosymplectic Hamiltonian systems}
\protect\label{kchs1}

Consider the $k$-cosymplectic manifold
$(\rk\times(T^1_k)^*Q,\eta^A,\Omega^A,{\cal V})$, and let ${\cal
H}\in\Cinfty( \rk\times(T^1_k)^*Q)$ be a Hamiltonian function.
$(\rk\times(T^1_k)^*Q,{\cal H})$ is called a {\sl $k$-cosymplectic
Hamiltonian system}. The {\sl HDW-equations} for this system are the
set of partial diferential equations:
 \begin{equation}
 \label{he}
  \frac{\partial {\cal H}}{\partial q^i}=-\ds\sum_{A=1}^k\frac{\partial\bar\psi^A_i}{\partial t^A}
 \quad , \quad
 \frac{\partial {\cal H}}{\partial p^A_i}=\frac{\partial\bar\psi^i}{\partial t^A}\quad ; \quad 1\leq A\leq k,
\, 1\leq\ i \leq n \, .
\end{equation}
where the solutions $\bar\psi(t)=(t,\bar\psi^i(t),\bar\psi^A_i(t))$
are sections of the projection $\bar\pi_k:\rk\times (T^1_k)^*Q \to
\rk$.

We denote by
$\vf^k_{\cal H}(\rk\times(T^1_k)^*Q)$ the set of $k$-vector fields
${\bf \bar X}=(\bar X_1,\dots,\bar X_k)$ on $\rk\times(T^1_k)^*Q$
wich are solutions to the equations
\begin{equation}
\label{geonah}
 \ds\sum_{A=1}^k \inn(\bar X_A)\Omega^A =\d {\cal
H}-\ds\sum_{A=1}^k R_A({\cal H}) \eta^A \quad , \quad \eta^A(\bar
X_B)=\delta^A_B\ ; \quad 1\leq A,B\leq k\ .
\end{equation}
Since $R_A=\partial/\partial t^A$ and $\eta^A=\d t^A$, then we can
write locally  the above equations as follows
\begin{equation}
\label{geonah1}
 \ds\sum_{A=1}^k\inn(\bar X_A)\Omega^A = \d {\cal H}-
\ds\sum_{A=1}^k\frac{\partial {\cal H}}{\partial t^A}\d t^A
\quad ,\quad \d t^A(\bar X_B)=\delta^A_B\quad ; \quad 1\leq
A,B\leq k\ .
\end{equation}

 In a local system of coordinates, $\bar X_A$ are locally given by
 \beq
\bar X_A=(\bar X_A)^B\frac{\partial}{\partial t^B} +(\bar X_A)^i\frac{\partial}{\partial q^i}
  +(\bar X_A)^B_i\frac{\partial}{\partial p^B_i} \ .
 \label{xcoor} 
\eeq
and, using (\ref{locexp0}),  we obtain that the equations (\ref{geonah})
 are equivalent to the equations
\begin{equation}
\label{111}
 \frac{\partial {\cal H}}{\partial p^A_i}=(\bar X_A)^i \quad ,\quad
\frac{\partial {\cal H}}{\partial q^i}= -\ds\sum_{A=1}^k(\bar
X_A)^A_i  \quad ,\quad (\bar X_A)^B=\delta_A^B \ ,
\end{equation}

The existence of $k$-vector fields that are solutions to (\ref{geonah1})
 is assured, and in a local system of coordinates they depend on
$n(k^2-1)$ arbitrary functions, but for integrable solutions
the number of arbitrary functions is, in general, less than $n(k^2-1)$.

\begin{proposition}
 Let ${\bf \bar X}=(\bar X_1,\dots,\bar X_k)$ be an integrable
 $k$-vector field in $\rk\times(T^1_k)^*Q$
and $\bar\psi\colon\Real^k\to\rk\times (T^1_k)^*Q$ an integral
section of ${\bf\bar X}$. Then
$\bar\psi(t)=(t,\bar\psi^i(t),\bar\psi^A_i(t))$ is a solution to the
HDW-equations (\ref{he}) if, and only if,
 ${\bf \bar X}\in\vf^k_{\cal H}(\rk\times(T^1_k)^*Q)$.
\label{prop2}
\end{proposition}
\proof
If $\bar\psi(t)=(t,\bar\psi^i(t),\bar\psi^A_i(t))$ is an
integral section of ${\bf \bar X}$, we have that
 \begin{equation}
\frac{\partial \bar\psi^i}{\partial t^B}=(\bar X_B)^i \quad ,
\quad \frac{\partial \bar\psi^A_i}{\partial t^B}=(\bar X_B)^A_i \, ,
\label{intsec}
 \end{equation}
and therefore we obtain that (\ref{111}) are the HDW-equations
(\ref{he20}). \qed

Furthermore we have:

\begin{proposition}
Every section $\bar\psi\colon\rk\to \rk\times(T^1_k)^*Q$ of the
projection $\bar\pi_k$ that is a solution  to the HDW-equations
(\ref{he}) is an integral section of an integrable $k$-vector field
 ${\bf \bar X}\in\vf^k_{\cal H}(\rk\times(T^1_k)^*Q)$.
\label{prop5}
\end{proposition}
\proof Let $\bar\psi\colon U_0\subset\rk\to\rk\times(T^1_k)^*Q$ be a
section of the projection $\bar\pi_k$ that is a solution  to the
HDW-equations (\ref{he}). We have that $\bar\psi$ is an injective
immersion and ${\rm Im}\,\bar\psi$ is a closed submanifold of
$\rk\times(T^1_k)^*Q$, since ${\rm Im}\,\bar\psi={\rm graph}\,\psi$,
for $\psi=\bar\pi_2\circ\bar\psi\colon\rk\to(T^1_k)^*Q$. Then the
construction of the integrable $k$-vector field in $\rk\times
(T^1_k)^*Q$, which has $\bar\psi$ as integral section and is a
solution to (\ref{geonah}), follows the same pattern as in
proposition \ref{prop3}.
 \qed

So the equations (\ref{geonah}) are a geometric version of the
HDW-equations(\ref{he}).

\subsection{Autonomous $k$-cosymplectic Hamiltonian systems}
\protect\label{akcs}

Following a terminology analogous to that in mechanics, we define:

\begin{definition}
A $k$-cosymplectic Hamiltonian system
 $(\rk\times(T^1_k)^*Q,{\cal H})$
is said to be {\rm autonomous} if \ $\ds\Lie(R_A){\cal
H}=\derpar{{\cal H}}{t^A}=0$, for $1\leq A\leq k$.
\label{autonomous}
\end{definition}

Observe that the condition in definition \ref{autonomous} means
that ${\cal H}$ does not depend on the variables $t^A$, and thus
${\cal H}=\bar\pi_2^*H$ for some $H\in\Cinfty((T^1_k)^*Q)$.

For an autonomous $k$-cosymplectic Hamiltonian system, the equations
(\ref{geonah}) become
\begin{equation}
\label{geonahaut}
 \ds\sum_{A=1}^k\inn(\bar X_A)\Omega^A =\d {\cal H} \quad , \quad
\eta^A(\bar X_B)=\delta^A_B\ ; \quad 1\leq A,B\leq k\ .
\end{equation}

Therefore:

\begin{proposition}
 Every autonomous $k$-cosymplectic Hamiltonian system
 $(\rk\times(T^1_k)^*Q,{\cal H})$
defines a $k$-symplectic Hamiltonian system $((T^1_k)^*Q,H)$, where
${\cal H}=\bar\pi_2^*H$, and conversely.
\end{proposition}

We have the following result for solutions to the Hamilton-de
Donder-Weyl equations:

\begin{theorem}
Let $(\rk\times(T^1_k)^*Q,{\cal H})$ be an autonomous
$k$-cosymplectic Hamiltonian system and let $((T^1_k)^*Q,H)$ be its
associated $k$-symplectic Hamiltonian system . Then, every section
$\bar\psi\colon\rk\to\rk\times(T^1_k)^*Q$, that is, a solution to the
HDW-equations (\ref{he}) for the system $(\rk\times(T^1_k)^*Q,{\cal
H})$ defines a map $\psi\colon\rk\to (T^1_k)^*Q$ that is a solution
to the HDW-equations (\ref{he20}) for the system $((T^1_k)^*Q,H)$;
and conversely. \label{onetoone}
\end{theorem}
\proof Since ${\cal H}=\bar\pi_2^*H$ we have
\begin{equation}\label{h}
\ds\frac{\partial{\cal H}}{\partial q^i}= \ds\frac{\partial
H}{\partial q^i}\quad,\quad \ds\frac{\partial \cal H}{\partial
p^A_i}= \ds\frac{\partial H}{\partial p^A_i}\;.
\end{equation}

Let $\bar\psi\colon\rk\to\rk\times(T^1_k)^*Q$ be a section of the
projection $\bar\pi_k$, which in coordinates is expressed as
$\bar\psi(t)=(t,\bar\psi^i(t),\bar\psi^A_i(t))$. Then we construct
the map $\psi=\bar\pi_2\circ\bar\psi\colon\rk\to(T^1_k)^*Q$, which
in coordinates is expressed as
$\psi(t)=(\psi^i(t),\psi^A_i(t))=(\bar\psi^i(t),\bar\psi^A_i(t))$.
Then, if $\bar\psi$ is a solution to the HDW-equations (\ref{he}),
from (\ref{h}) we obtain that $\psi$ is a solution to the
HDW-equations (\ref{he20}).

Conversely, consider a map $\psi\colon\rk\to(T^1_k)^*Q$. We define
$\bar\psi=(Id_{\Real^k},\psi):\rk\to\rk\times(T^1_k)^*Q$.
Furthermore, if $\psi(t)=(\psi^i(t),\psi^A_i(t))$, then
$\bar\psi(t)=(t,\bar\psi^i(t),\bar\psi^A_i(t))$, with
$\bar\psi^i(t)=\psi^i(t)$ and $\bar\psi^A_i(t)=\psi^A_i(t)$ (observe
that, in fact,  ${\rm Im}\,\bar\psi={\rm graph}\,\psi$). Hence,  if
$\psi$ is a solution to the HDW-equations (\ref{he20}), from
(\ref{h}) we obtain that $\bar\psi$ is a solution to the
HDW-equations (\ref{he}). \qed

For $k$-vector fields that are solutions to the geometric field
equations (\ref{generic0}) and (\ref{geonahaut}) we have:

\begin{proposition}
Let $(\rk\times(T^1_k)^*Q,{\cal H})$ be an autonomous
$k$-cosymplectic Hamiltonian system and let $((T^1_k)^*Q,H)$ be its
associated $k$-symplectic Hamiltonian system. Then every $k$-vector
field ${\bf X}\in\vf^k_H(T^1_k)^*Q)$ defines a  $k$-vector field
${\bf \bar X}\in\vf^k_{\cal H}(\rk\times(T^1_k)^*Q)$.

Furthermore, ${\bf X}$ is integrable if, and only if, its associated
${\bf \bar X}$ is integrable too.
\end{proposition}
\proof
 Let ${\bf X}=(X_1,\dots,X_k)\in\vf^k_H((T^1_k)^*Q)$. For
every $A=1,\ldots,k$, let $\bar X_A\in\vf(\rk\times(T^1_k)^*Q)$ be
the {\sl suspension} of the corresponding vector field
$X^A\in\vf((T^1_k)^*Q)$, which is defined as follows (see \cite{am},
p. 374, for this construction in mechanics): for every ${\rm
p}\in(T^1_k)^*Q$, let $\gamma^A_{\rm p}\colon\r\to(T^1_k)^*Q$ be the
integral curve of $X_A$ passing through ${\rm p}$; then, if
$t_0=(t_0^1,\ldots,t_0^k)\in\rk$, we can construct the curve
$\bar\gamma^A_{\bar{\rm p}}\colon\r\to\rk\times(T^1_k)^*Q$, passing
through the point $\bar{\rm p}\equiv(t_0,{\rm
p})\in\rk\times(T^1_k)^*Q$, given by $\bar\gamma^A_{\bar{\rm
p}}(t^A)=(t_0^1,\ldots,t^A+t_0^A,\ldots,t_0^k;\gamma_{\rm p}(t^A))$.
Therefore, $\bar X_A$ is the vector field tangent to
$\bar\gamma^A_{\bar{\rm p}}$ at $(t_0,{\rm p})$. In natural
coordinates, if $X_A$ is locally given by (\ref{coorxa}), then $\bar
X_A$ is locally given by
$$
\bar X_A =\derpar{}{t^A}+(\bar X_A)^i\frac{\partial}{\partial q^i}+
(\bar X_A)_i^B\frac{\partial}{\partial p_i^B}=
\derpar{}{t^A}+\bar\pi_2^*(X_A)^i\frac{\partial}{\partial  q^i}+
\bar\pi_2^*(X_A)_i^B\frac{\partial}{\partial p_i^B}\, .
$$
Observe that $\bar X_A$ are $\bar\pi_2$-projectable vector fields,
and $(\bar\pi_2)_*\bar X_A=X_A$. In this way we have defined a
$k$-vector field ${\bf\bar  X}=(\bar X_1,\dots,\bar X_k)$ in
$\rk\times(T^1_k)^*Q$. Therefore, taking (\ref{symcosym}) into
account,
$$
\ds\sum_{A=1}^k\inn(\bar X_A)\Omega^A-\d{\cal
H}=\ds\sum_{A=1}^k\inn(\bar
X_A)\bar\pi_2^*\omega^A-\d(\bar\pi_2^*H)=
\bar\pi_2^*(\sum_{A=1}^k\inn((\pi_2)_*\bar X_A)\omega^A-\d H)=0 \ ,
$$
since ${\bf X}=(X_1,\dots,X_k)\in\vf^k_H(T^1_k)^*Q)$, and therefore
${\bar  X}=(\bar X_1,\dots,\bar X_k)\in\vf^k_{\cal
H}(\rk\times(T^1_k)^*Q)$.

Furthermore, if $\psi\colon\rk\to (T^1_k)^*Q$ is an integral section
of ${\bf X}$, then  $\bar\psi\colon\rk\to\rk\times(T^1_k)^*Q$ such
that $\bar\psi=(Id_{\rk},\psi)$ (see Theorem \ref{onetoone}) is an
integral section of ${\bf \bar X}$.

Now, if $\bar\psi$ is an integral section of ${\bf \bar X}$, the
equations (\ref{intsec}) hold for
$\bar\psi(t)=(t,\bar\psi^i(t),\bar\psi^A_i(t))$ and, as $(\bar
X_A)^i=\bar\pi_2^*(X_A)^i$ and $(\bar
X_A)_i^B=\bar\pi_2^*(X_A)_i^B$, this is equivalent to saying that
the equations (\ref{intsec0}) hold for
$\psi(t)=(\psi^i(t),\psi^A_i(t))$; that is, $\psi$ is an integral
section of ${\bf X}$. \qed

{\bf Remark}: The converse statement is not true. In fact, the
$k$-vector fields that are solution to the geometric field equations
(\ref{geonahaut}) are not completely determined, as the equations
(\ref{111}) show, and then there are $k$-vector fields in
$\vf^k_{\cal H}(\rk\times(T^1_k)^*Q)$ that are not
$\bar\pi_2$-projectable (in fact, it suffices to take their
undetermined component functions to be not $\bar\pi_2$-projectable).
However, we have the following particular result:

\begin{proposition}
Let $((T^1_k)^*Q,H)$ be an admissible $k$-symplectic Hamiltonian
system, and $(\rk\times(T^1_k)^*Q,{\cal H})$ its associated
autonomous $k$-cosymplectic Hamiltonian system. Then, every
integrable $k$-vector field ${\bf \bar X}\in\vf^k_{\cal
H}(\rk\times(T^1_k)^*Q)$ defines an integrable $k$-vector field
${\bf X}\in\vf^k_H((T^1_k)^*Q)$.
\end{proposition}
\proof
 If ${\bf \bar X}\in\vf^k_{\cal H}(\rk\times(T^1_k)^*Q)$ is
an integrable $k$-vector field, denote by $\bar{\cal S}$ the set of
its integral sections (i.e., solutions to the the HDW-equations
(\ref{he})). Let ${\cal S}$ be the set of maps
$\psi\colon\rk\to(T^1_k)^*Q$ associated with these sections by
Theorem \ref{onetoone}, which are admissible solutions to the
HDW-equations (\ref{he20}), by the hypothesis that
$((T^1_k)^*Q,\omega^A,H)$ is an admissible $k$-symplectic
Hamiltonian system. Then, by proposition \ref{prop3} we can
construct an integrable $k$-vector field ${\bf
X}\in\vf^k_H((T^1_k)^*Q)$ for which ${\cal S}$ is its set of
integral sections (which are admissible solutions to the
HDW-equations (\ref{he20})).
 \qed

\section{$k$-symplectic and $k$-cosymplectic Lagrangian formalisms}

(See \cite{fam,RRS} for details on the construction of this
formalism).

\subsection{Canonical structures in the bundles $T^1_kQ$
and $\rk \times T^1_kQ$}

Consider the bundle $\tau^1_Q\colon T^1_kQ\to Q$
(see Section \ref{kvfis}).
If $(q^i)$ are local coordinates on $U \subseteq Q$ then the induced
local coordinates $(q^i , v^i)$, $1\leq i \leq n$, in
$TU=(\tau^1_Q)^{-1}(U)$ are given by $q^i(v_q)=q^i(q)$,
$v^i(v_q)=v_q(q^i)$, and  the induced local coordinates $(q^i ,
v_A^i)$, $1\leq i \leq n,\, 1\leq A \leq k$, in
$T^1_kU=(\tau^1_Q)^{-1}(U)$ are given by
$$
q^i({v_1}_q,\ldots , {v_k}_q)=q^i(q),\qquad v_A^i({v_1}_q,\ldots ,
{v_k}_q)={v_A}_q(q^i) \, .
$$

For a vector $Z_q\in T_qQ$, and for $A=1,\ldots, k$, we define its
{\sl vertical $A$-lift}, $(Z_q)^{V_A}$, at the point
$({v_1}_q,\ldots,{v_k}_q)\in T_k^1Q$, as the vector tangent to the
fiber $(\tau^1_Q)^{-1}(q)\subset T_k^1Q$, which is given by
  $$
(Z_q)^{V_A}({v_1}_q,\ldots,v_A)=
\frac{d}{ds}({v_1}_q,\ldots,{v_{A-1}}_q,v_{A_q}+sZ_q,{v_{A+1}}_q,\ldots,{v_k}_q)\vert_{s=0} \;.
 $$
 In  local coordinates, if $X_q = a^i \,\ds\frac{\partial}{\partial q^i}\Big\vert_q$, we have
$\ds  (Z_q)^{V_A}({v_1}_q,\ldots,{v_k}_q)= a^i
\displaystyle\frac{\partial}{\partial
v^i_A}\Big\vert_{({v_1}_q,\ldots,{v_k}_q)}$.
Then, the {\sl canonical $k$-tangent structure} on $T^1_kQ$ is the set
$(S^1,\ldots,S^k)$ of tensor fields  of type $(1,1)$ defined by
 $$
S^A(w_q)(Z_{w_q})= ((\tau^1_Q)_*(w_q)(Z_{w_q}))^{V_A}(w_q) \quad , \quad
\mbox{\rm for $w_q\in T^1_kQ$, $Z_{w_q}\in T_{w_q}(T^1_kQ)$; $A=1,
\ldots , k$}\,.
 $$
 In local  coordinates  we have
 \beq
\label{localJA} 
S^A=\frac{\partial}{\partial v^i_A} \otimes \d q^i\;.
 \eeq
The {\sl Liouville vector field}
$\Delta\in\vf(T^1_kQ)$  is the infinitesimal generator of the following flow
 $$
\psi\colon\Real\times T^1_kQ\longrightarrow T^1_kQ  \quad , \quad
  \psi(s,v_{1_{q}}, \ldots ,v_{k_{q}})= (e^s   v_{1_{q}},\ldots , e^s   v_{k_{q}})\, ,
$$
and in local coordinates it has the form
$$
 \Delta =\sum_{A=1}^kv^i_A \derpar{}{v_A^i}\ .
$$

Now, consider the manifold $J^1\pi_{\rk}$ of 1-jets of sections
 of the trivial bundle $\pi_{\rk}\colon \rk \times Q \to \rk$, which is
 diffeomorphic to $\rk \times  T^1_kQ$, via the
diffeomorphism given by
\begin{equation} \label{isomor}
\begin{array}{rcl}
J^1\pi_{\rk} & \to & \rk   \times T^1_kQ \\
\noalign{\medskip} j^1_t\phi= j^1_t(Id_{\rk},\phi_Q) & \to & (
t,v_1, \ldots ,v_k) \ ,
\end{array}
\end{equation}
where $\phi_Q\colon\rk \stackrel{\phi}{\to}  \rkq \stackrel{\pi_Q}{\to}Q$,  and
$\displaystyle v_A=(\phi_Q)_*(t)(\ds\frac{\partial}{\partial t^A}\Big\vert_t)$, for $1\leq A \leq k$.
We denote by $\bar\tau^1_Q\colon\rk\times  T^{1}_{k} Q \to Q$ the canonical projection.
 If $(q^i)$ are local coordinates on $U \subseteq Q$,  then the induced
local coordinates  $(t^A,q^i , v^i_A)$ on $(\bar\tau^1_Q)^{-1}(U)=\rk \times T^1_kU$ are
$$
t^A(t,{v_1}_q,\ldots , {v_k}_q)   =   t^A; \quad
q^i(t,{v_1}_q,\ldots , {v_k}_q) =q^i(q); \quad
v_A^i(t,{v_1}_q,\ldots , {v_k}_q)   = {v_A}_q(q^i)\, .
 $$

We consider the extension of $S^A$ to $\rk\times
T^1_kQ$, which we denote  by $\bar S^A$, and they have the same
local expressions (\ref{localJA}).
Finally, we introduce the {\sl Liouville vector field} $\bar\Delta\in\vf(\rk\times T^1_kQ)$,
which is the infinitesimal generator of the following flow
$$
\begin{array}{ccc}
\r \times (\r^{k}\times T^1_kQ) & \longrightarrow & \r^{k}\times
T^1_kQ  \\ \noalign{\medskip} (s,(t,{v_1}_q,\ldots , {v_k}_q)) &
\longrightarrow & (t, e^s{v_1}_q, \ldots,e^s{v_k}_q)\, ,
\end{array}
$$
  and in local coordinates it has the form
\begin{equation}\label{locci}
\bar \Delta=   \displaystyle\sum_{i,A} v^i_A
\frac{\displaystyle\partial}{\displaystyle\partial v^i_A}\, ,
\end{equation}

\subsection{$k$-symplectic Lagrangian formalism}
\protect\label{kslf}

  Let $L\in\Cinfty(T^1_kQ)$ be a Lagrangian function.

A family of forms  $\theta_L^A\in\df^1(T^1_kQ)$, $1\leq A \leq k$,
is introduced by using the $k$-tangent structure of $T^1_kQ$, as follows
$$
 \theta_L^A=  \d L \circ S^A  \, \quad 1 \leq A \leq k  \quad ,
$$
and hence we define $\omega_L^A=-\d\theta_L^A$.
In coordinates
$$
 \theta_L^A=\frac{\partial L}{\partial v^i_A}\, \d q^i
 \quad ,\quad
\omega_L^A =
\d q^i \wedge \d\left(\frac{\partial L}{\partial v^i_A}\right)=
\frac{\partial ^2 L}{\partial q^j\partial v^i_A}\d q^i\wedge\d q^j +
\frac{\partial ^2 L}{\partial v^j_B\partial v^i_A}\d q^i\wedge\d v^j_B \ .
$$
We can also define the {\sl Energy Lagrangian function} associated to $L$,
$E_L\in\Cinfty(T^1_kQ)$,  as $\ds E_L=\Delta (L) -L$.
 Its local expression is
$$
E_L=v^i_A\frac{\partial L}{\partial v^i_A}-L \ .
$$
Finally, the  Legendre map  $ FL\colon T^1_kQ \longrightarrow (T^1_k)^*Q$
  was introduced by G\"{u}nther \cite{gun}, and
 we rewrite  it  as follows: if $(v_{1_q}, \dots ,v_{k_q}) \in (T^1_k)_qQ$
$$
[FL(v_{1_q}, \dots , v_{k_q})]^A(w_q)=
\displaystyle\frac{d}{ds}\displaystyle L(v_{1_q}, \dots
,v_{A_q}+sw_q, \ldots , v_{k_q})\vert_{s=0}  ,
$$
 for each $A=1, \ldots, k$. We have that  $FL$ is locally given by
 \begin{equation}
 \label{locfl}
(q^i,v^i_A)  \longrightarrow \left(q^i, \frac{\displaystyle\partial
L}{\displaystyle\partial v^i_A}\right).
\end{equation}
 Furthermore, from (\ref{locexp0}) and (\ref{locfl}) we obtain that
 \begin{equation}
 \label{mm}
\theta_L^A=FL^*\theta^A\quad,\quad \omega_L^A=FL^*\omega^A
 \end{equation}

The Lagrangian $L$ is said to be {\sl regular} if
 $(\frac{\partial^2 L}{\partial v^i_A \partial v^j_B})$ is a non-singular matrix
at every point of $T^1_kQ$.
Then,  from (\ref{locfl}) and  (\ref{mm}) we get:

\begin{proposition}
 Let $L\in\Cinfty(T^1_kQ)$ be a Lagrangian. The following conditions are equivalent:

1) $L$ is regular. 2) $FL$ is a local diffeomorphism. 3)
$(T^1_kQ,\omega_L^A,V)$, where $V=Ker (\tau^1_Q)_*$, is a
$k$-symplectic manifold.
\qed
 \end{proposition}

A Lagrangian function $L$
is said to be {\sl hyperregular} if the corresponding
Legendre map  $FL$ is a global diffeomorphism.
 If $L$ is regular, $(T^1_kQ,L)$ is said to be a
{\sl $k$-symplectic Lagrangian system}. If $L$ is not
regular $(T^1_kQ,L)$ is a {\sl $k$-presymplectic Lagrangian system}.

The {\sl Euler-Lagrange equations} for $L$ are:
\begin{equation}
\label{ELe}
 \sum_{A=1}^k\frac{\partial}{\partial t^A}\Big\vert_t
\left(\frac{\displaystyle\partial L}{\partial
v^i_A}\Big\vert_{\varphi(t)}\right)= \frac{\partial L}{\partial
q^i}\Big\vert_{\varphi(t)}
 \quad , \quad
v^i_A(\varphi(t))= \frac{\partial\varphi^i}{\partial t^A} \quad ,
\quad 1\leq i \leq n, \,  1\leq A \leq k
\end{equation}
whose solutions are maps $\varphi\colon\Real^k \to T^1_kQ$ that, as
a consequence of the last group of equations (\ref{ELe}),
 are first prolongations to $T^1_kQ$
 of maps $\phi=\tau^1_Q\circ\varphi\colon\Real^k \to Q$;
that is, $\varphi$ are {\sl holonomic}. This means that
$\varphi=\phi^{(1)}$ where
$$\begin{array}{rlcl}
\phi^{(1)}:&\Real^k &\to& T^1_kQ\\\noalign{\medskip} & t&\mapsto
&\phi^{(1)}(t)=(\phi_*(t)\big(\frac{\partial}{\partial
t^1}\Big\vert_{t}\big),\ldots,\phi_*(t)\big(\frac{\partial}{\partial
t^k}\Big\vert_{t}\big))
\end{array}\quad .$$

Let $\vf^k_L(T^1_kQ)$ be the set of $k$-vector fields ${\bf
\Gamma}=(\Gamma_1,\dots,\Gamma_k)$ in $T^1_kQ$, wich are solutions to
 \begin{equation}
\label{genericEL}
 \sum_{A=1}^k \inn(\Gamma_A)\omega_L^A=\d E_L\, .
 \end{equation}
If $\ds\Gamma _A  =  ( \Gamma _A)^i \frac{\partial}{\partial  q^i} +
( \Gamma _A)^i_B\frac{\partial}{\partial v^i_B}$ locally,
 then ${\bf \Gamma}$ is a solution to (\ref{genericEL})
 if, and only if, $( \Gamma_A)^i$ and $( \Gamma _A)^i_B$ satisfy
\beann
  \left( \frac{\partial^2 L}{\partial q^i \partial v^j_A} + \frac{\partial^2 L}{\partial q^j \partial v^i_A}
\right) \, ( \Gamma _A)^j - \frac{\partial^2 L}{\partial v_A^i
\partial v^j_B} \, ( \Gamma _A)^j_B &=&
 v_A^j \frac{\partial^2 L}{\partial q^i\partial v^j_A} - \frac{\partial  L}{\partial q^i }
\\
\frac{\partial^2 L}{\partial v^j_B\partial v^i_A} \, (\Gamma_A)^i
 &=& \frac{\partial^2 L}{\partial v^j_B\partial v^i_A} \, v_A^i \, .
\eeann
If the Lagrangian is regular, the above equations  are equivalent to
$$
\frac{\partial^2 L}{\partial q^j \partial v^i_A} v^j_A +
\frac{\partial^2 L}{\partial v_A^i\partial v^j_B}( \Gamma _A)^j_B =
\frac{\partial  L}{\partial q^i} \quad , \quad
 ( \Gamma _A)^i = v_A^i \ .
$$
The last group of these equations is the local expression of the
condition that ${\bf \Gamma}$ is a {\sc sopde} (see \cite{fam}), and hence, if it is
integrable, its integral sections are first prolongations
 $\phi^{(1)}\colon\Real^k\to T^1_kQ$ of
maps $\phi\colon \Real^k \to Q$, and using the first group of
equations, we deduce that $\phi^{(1)}$ are solutions to the
Euler-Lagrange equations (\ref{ELe}). If $L$ is not regular then, in
general, the equations (\ref{ELe}) or (\ref{genericEL}) have no
solutions anywhere in $T^1_kQ$, but they do in a submanifold $S$ of $T^1_kQ$
(in the most favourable situations). Moreover, solutions to
(\ref{genericEL}) are not {\sc sopde} necessarily.

We define {\sl admissible solutions} to the Euler-Lagrange equations
and {\sl admissible $k$-symplectic Lagrangian systems} in the same
way as in the Hamiltonian case (definition \ref{hyp}). Then the
statement of Proposition \ref{prop3} can be proved analogously for
these admissible solutions.
This proof holds for regular $k$-symplectic
Lagrangian systems, and for the non-regular case the proof is still valid
considering the submanifold $S$ of $(T^1_k)^*Q$ where
the Lagrangian field equations have solutions.

\subsection{$k$-cosymplectic Lagrangian formalism and
 autonomous $k$-cosymplectic Lagrangian systems}
\protect\label{kcolag}

Let ${\cal L}\in\Cinfty(\rk \times T^1_kQ)$ be a Lagrangian .

A family of forms  $\Theta_{\Lag}^A\in\df^1(\rk\times T^1_kQ)$, $1\leq A \leq k$,
is introduced by using the $k$-tangent structure of $\rk\times T^1_kQ$, as follows
$$
 \Theta_{\Lag}^A=  \d{\cal L}\circ \bar S^A  \, \quad 1 \leq A \leq k  \quad ,
$$
and hence we define $\Omega_{\Lag}^A=-\d\Theta_{\Lag}^A$.
In coordinates
\begin{equation}\label{am2}
 \Theta_{\Lag}^A=\frac{\partial\Lag}{\partial v^i_A}\, \d q^i
 \quad ,\quad
 \Omega_{\Lag} ^A =
 \frac{\partial ^2 \Lag }{\partial q^j\partial v^i_A}\d q^i\wedge\d q^j +
 \frac{\partial ^2 \Lag }{\partial v^j_B\partial v^i_A}\d q^i\wedge\d v^j_B +
 \frac{\partial ^2 \Lag }{\partial t^B\partial v^i_A}\d q^i\wedge\d t^B  \ .
\end{equation}
We can also define the {\sl Energy Lagrangian function} associated to $\Lag$,
${\cal E}_{\Lag} \in\Cinfty(\rk\times T^1_kQ)$ as
$\ds {\cal E}_{\Lag}=\bar\Delta (\Lag) -\Lag$,
whose local expression is
$$
 {\cal E}_{\Lag} =v^i_A\frac{\partial \Lag}{\partial v^i_A}-\Lag\ .
$$
Finally,  the  Legendre map
$F{\cal L}\colon\rk\times T^1_kQ \longrightarrow \rk\times (T^1_k)^*Q$,
 is defined  as follows:
 $$
F{\cal L}(t,{v_1}_q,\ldots , {v_k}_q)=(t,\ldots, [F{\cal L}(t,{v_1}_q,\ldots , {v_k}_q)]^A,\ldots )
 $$
 where
$$
 [F{\cal L}(t,{v_1}_q,\ldots , {v_k}_q)]^A(w_q)=
\displaystyle\frac{d}{ds}\displaystyle {\cal L}\left( t,{v_1}_q,
\dots ,{v_A}_q+sw_q, \ldots , {v_k}_q) \right)\vert_{s=0} ,
$$
 for each $A=1, \ldots, k$; and it is locally given by
\begin{equation}\label{locfl1}
F{\cal L}:(t^A,q^i,v^i_A)  \longrightarrow  \left(t^A,q^i,
\frac{\displaystyle\partial {\cal L}}{\displaystyle\partial v^i_A }\right)\, .
\end{equation}
It is obvious that
  \begin{equation}
\label{am}
\Theta_{\cal L}^A=F{\cal L}^*\Theta^A\, , \quad\Omega_{\cal
L}^A=F{\cal L}^*\Omega^A, \quad 1\leq A\leq k \quad .
\end{equation}
Observe that
$F{\Lag}={\rm Id}_{_{\rk}}\times FL\colon\rk\times
T^1_kQ\to\rk\times(T^1_k)^*Q$, (see (\ref{symcosym}), (\ref{mm}) and (\ref{am})).

The Lagrangian ${\cal L}={\cal L}(t^B,q^j,v^j_B)$ is {\sl regular} if the matrix
 $(\frac{\partial^2 \Lag }{\partial v^i_A \partial v^j_B})$ is not singular
at every point of $\rk\times T^1_kQ$.
Then, from (\ref{lockcos}), (\ref{locfl1}) and (\ref{am}) we deduce the
following proposition (See \cite{mod2}):

\begin{proposition}
 Let ${\Lag}\in\Cinfty(\rk \times T^1_kQ)$ be a Lagrangian. The following conditions are equivalent:

1) $\Lag$ is regular. 2) $F\Lag$ is a local diffeomorphism. 3)
$(\rk\times T^1_kQ,\d t^A,\Omega_{\Lag}^A,{\cal V})$, where ${\cal V}=\ker\, (\bar\tau_0)_*$,
 is a $k$-cosymplectic manifold.
 \end{proposition}
\begin{flushright}
\qed
 \end{flushright}

A Lagrangian function ${\Lag}$
is said to be {\sl hyperregular} if the corresponding
Legendre map  $F\Lag$ is a global diffeomorphism.
 If $\Lag$ is regular, $(\rk\times T^1_kQ,\Lag)$ is said to be a
{\sl $k$-cosymplectic Lagrangian system}.
If $\Lag $ is not regular,  $(\rk\times T^1_kQ,\Lag)$ is a
 {\sl $k$-precosymplectic Lagrangian system}.

The {\sl Euler-Lagrange equations} are (\ref{ELe}), but now the
Lagrangian is ${\Lag}={\Lag}(t^B,q^j,v^j_B)$, and their solutions
are sections $\bar\varphi\colon\Real^k \to\rk\times T^1_kQ$ of the
natural projection $\rk\times T^1_kQ\to\rk$, which are first
prolongations to $\rk\times T^1_kQ$ of sections
 $\phi\colon\Real^k \to Q$ of the natural projection  $\rk\times Q\to\rk$;
that is, $\bar\varphi$ are {\sl holonomic}. This means that
$\bar\varphi=\phi^{[1]}$ where
$$
\begin{array}{rcl}
\phi^{[1]}:\r^k & \longrightarrow &   \r^k \times T^1_kQ \\ t &
\longrightarrow &
\phi^{[1]}(t)=\left(t,\phi_*(t)(\ds\frac{\partial}{\partial t^1}),
\ldots , \phi_*(t)(\ds\frac{\partial}{\partial t^k}) \right)
\end{array}
$$
 Furthermore, we denote
by $\vf^k_{\Lag} (\rk\times T^1_kQ)$ the set of $k$-vector fields
${\bf \bar\Gamma}=(\bar\Gamma_1,\dots,\bar\Gamma_k)$ in $\rk\times
T^1_kQ$,
 that are solutions to the equations
 \begin{equation}
\label{genericELe}
 \ds\sum_{A=1}^k\inn(\bar\Gamma_A)\Omega^A_{\Lag} = \d {\cal E}_{\Lag}-
\ds\sum_{A=1}^k\frac{\partial {\cal L}}{\partial t^A}\d t^A \quad
,\quad \d t^A(\bar\Gamma_B)=\delta^A_B\quad ; \quad 1\leq A,B\leq k\
.
 \end{equation}
In a local system of natural coordinates, if
 \beq
\bar\Gamma _A  =  (\bar\Gamma _A)^B \frac{\partial}{\partial t^B}+
 (\bar\Gamma _A)^i \frac{\partial}{\partial  q^i} +
 (\bar\Gamma _A)^i_B\frac{\partial}{\partial v^i_B}
\label{xlagcoor}
\eeq
 then ${\bf \bar\Gamma}$ is a solution to (\ref{genericELe})
 if, and only if,
$(\bar\Gamma_A)^i$ and $(\bar\Gamma _A)^i_B$ satisfy
 \bea
 (\bar\Gamma_A)^B=\delta_A^B \quad , \quad
(\bar\Gamma_A)^i\frac{\partial^2 L}{\partial t^B \partial v^i_A}=
 v^i_A\frac{\partial^2 \Lag}{\partial t^B\partial v^i_A} \quad , \quad
 (\bar\Gamma_A)^i\frac{\partial^2 \Lag}{\partial v^j_B \partial v^i_A}=
 v^i_A\frac{\partial^2 \Lag}{\partial v^j_B \partial v^i_A}
\nonumber \\
 \frac{\partial^2 \Lag}{\partial q^j \partial v^i_A}\left(v^i_A
 -(\bar\Gamma_A)^i\right)+\frac{\partial^2 \Lag}{\partial t^A \partial v^i_A}+
 v^k_A\frac{\partial^2 \Lag}{\partial q^k \partial v^i_A}+
 (\bar\Gamma_A)_B^k\frac{\partial^2 \Lag}{\partial v^k_B \partial v^i_A}=
 \frac{\partial  \Lag}{\partial q^i}
\label{elcoor} \eea When $\Lag$ is regular, we obtain that
$(\bar\Gamma_A)^i=v^i_A$, and the last equation can be written as
follows
\begin{equation}
 \frac{\partial^2 \Lag}{\partial t^A \partial v^i_A}+v^k_A
 \frac{\partial^2 \Lag}{\partial q^k \partial v^i_A}+ (\bar\Gamma_A)_B^k
 \frac{\partial^2 \Lag}{\partial v^k_B \partial v^i_A}=
 \frac{\partial  \Lag}{\partial q^i} \, ,
\label{elcoor1}
\end{equation}
then ${\bf \bar\Gamma}$ is a {\sc sopde} (see \cite{mod2}), and hence, if it is
integrable, its integral sections are holonomic and
 they are solutions to the Euler-Lagrange equations for $\Lag$.
If $\Lag $ is not regular, the existence of solutions to the
equations (\ref{ELe}) for $\Lag$ or to (\ref{genericELe}) is not
assured, in general, except in a submanifold of $T^1_kQ$ (in the
most favourable situations). Moreover, solutions to
(\ref{genericELe}) are not {\sc sopde} necessarily.

\begin{definition}
A $k$-cosymplectic (or  $k$-precosymplectic) Lagrangian system
is said to be {\rm autonomous} if \ $\ds \derpar{\cal L}{t^A}=0$ or,
what is equivalent, $\ds \derpar{{\cal E}_{\Lag}}{t^A}=0$, $1\leq
A\leq k$. \label{autonomousL}
\end{definition}

Now, all the results obtained in Section \ref{akcs} can be stated
and proved in the same way, considering the systems
 $(\rk\times T^1_kQ,\Lag)$
and $(T^1_kQ,L)$ instead of $(\rk\times (T^1_k)^*Q, {\cal H})$ and
$((T^1_k)^*Q,H)$.

Finally, the $k$-symplectic and $k$-cosymplectic Lagrangian and
Hamiltonian systems are related by means of the
 {\sl Legendre maps} $FL$ and  $F{\Lag}$.

\section{Multisymplectic Hamiltonian formalism}

\subsection{Multisymplectic manifolds and multimomentum bundles}
\protect\label{mmb}

(See, for instance, \cite{EMR-00}).

\begin{definition}
The couple $({\cal M},\Omega)$, with $\Omega\in\Omega^{k+1}({\cal
M})$ ($2\leq k+1\leq\dim\,{\cal M}$), is a {\rm multisymplectic
manifold}
 if $\Omega$ is closed and $1$-nondegenerate;
that is, for every $p\in{\cal M}$, and $X_p\in T_p{\cal M}$, we have
that $\inn(X_p)\Omega_p=0$ if, and only if, $X_p=0$.
\end{definition}

A very important example of multisymplectic manifold is the {\sl
multicotangent bundle} $\Lambda^k\Tan^*Q$ of a manifold $Q$, which
is the bundle of $k$-forms in $Q$, and is endowed with a canonical
multisymplectic  $(k+1)$-form.
 Other examples of multisymplectic manifolds which are relevant in
field theory are the so-called {\sl multimomentum bundles}: let
$\pi\colon E\to M$ be a fiber bundle, ($\dim\, M=k$, $\dim\,
E=n+k$), where $M$ is an oriented manifold with volume form
 $\omega\in\Omega^k(M)$, and denote by $(t^A,q^i)$
 ($1\leq A\leq k$, $1\leq n$) the natural coordinates in $E$
 adapted to the bundle, such that
 $\omega=\d t^1\wedge\ldots\wedge\d t^k\equiv{\rm d}^kt$.
First we have $\Lambda_2^kT^*E\equiv{\cal M}\pi$, which is the
bundle of $k$-forms on $E$ vanishing by the action of two
$\pi$-vertical vector fields. This is called the {\sl extended
multimomentum bundle}, and its canonical submersions are denoted by
$$
\kappa\colon{\cal M}\pi\to E \quad ; \quad
\bar\kappa=\pi\circ\kappa\colon{\cal M}\pi\to M
$$
We can introduce natural coordinates in ${\cal M}\pi$ adapted to the
bundle $\pi\colon E\to M$, which are denoted by $(t^A,q^i,p^A_i,p)$,
and such that $\omega={\rm d}^kt$. Then, denoting
$\ds\d^{k-1}t^A=\inn\left(\frac{\partial }{\partial t^A}\right){\rm
d}^kt$,  the elements of ${\cal M}\pi$ can be written as
   $p^A_i \, \d q^i \wedge d^{k-1}t_A \, + \,p\,\d^kt$.

${\cal M}\pi$ is a subbundle of $\Lambda^kT^*E$, and hence ${\cal
M}\pi$ is also endowed with canonical forms. First we have the
``tautological form'' $\Theta\in\Omega^k({\cal M}\pi)$, which is
defined as follows: let $(x,\alpha )\in\Lambda_2^kT^*E $, with $x\in
E$ and $\alpha\in\Lambda_2^kT_x^*E$; then, for every
$X_1,\ldots,X_m\in T_{(x,\alpha)}({\cal M}\pi)$, we have
\begin{equation}\label{t0}
\Theta ((x,\alpha ))(X_1,\ldots,X_m):= \alpha
(x)(T_{(x,\alpha)}\kappa(X_1),\ldots ,T_{(x,\alpha)}\kappa(X_m))
\end{equation}
Thus we define the multisymplectic form
 \begin{equation}\label{0t}
\Omega:=-{\rm d}\Theta\in\Omega^{k+1}({\cal M}\pi)
\end{equation}
and  the local expressions of the above forms are
\begin{equation}
 \Theta=p^A_i{\rm d} q^i\wedge{\rm d}^{k-1}t_A+p\,{\rm d}^kt
 \ , \
 \Omega=
-{\rm d} p^A_i\wedge{\rm d} q^i\wedge{\rm d}^{k-1}t_A-{\rm d}
p\wedge{\rm d}^kt \label{coormult}
\end{equation}
Consider $\pi^*\Lambda^kT^*M$, which is another bundle over $E$,
whose sections are the $\pi$-semibasic $k$-forms on $E$, and denote
by $J^1\pi^*$ the quotient $\Lambda_2^kT^*E/\pi^*\Lambda^kT^*M$.
$J^1\pi^*$ is usually called the {\sl restricted multimomentum
bundle} associated with the bundle $\pi\colon E\to M$. Natural
coordinates in $J^1\pi^*$ (adapted to the bundle $\pi\colon E\to M$)
are denoted by $(t^A,q^i,p^A_i)$.
 We have the natural submersions specified in the following diagram
$$
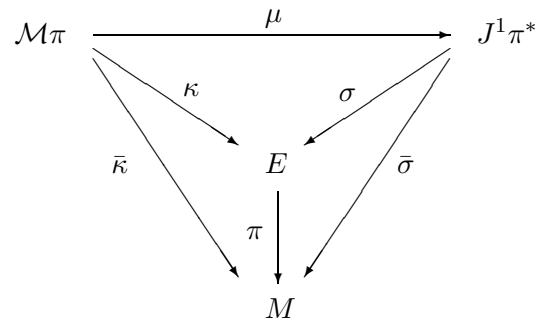
\begin{array}{ccc}
{\cal M}\pi &
\begin{picture}(135,20)(0,0)
\put(65,8){\mbox{$\mu$}} \put(0,3){\vector(1,0){135}}
\end{picture}
& J^1\pi^*
\\ &
\begin{picture}(135,100)(0,0)
\put(34,84){\mbox{$\kappa$}} \put(93,82){\mbox{$\sigma$}}
\put(7,55){\mbox{$\bar\kappa$}}
 \put(115,55){\mbox{$\bar\sigma$}}
 \put(58,30){\mbox{$\pi$}}
\put(65,55){\mbox{$E$}}
 \put(65,0){\mbox{$M$}}
\put(0,102){\vector(3,-2){55}}
 \put(135,102){\vector(-3,-2){55}}
\put(0,98){\vector(2,-3){55}}
 \put(135,98){\vector(-2,-3){55}}
\put(70,48){\vector(0,-1){35}}
\end{picture} &
\end{array}
$$

\subsection{Multisymplectic Hamiltonian formalism}
\protect\label{hsjpistar}

The Hamiltonian formalism in $J^1\pi^*$
presented here is based on the construction made in \cite{CCI-91}
(see also \cite{ELMR-2005} and \cite{EMR-00}).

\begin{definition}
 A section $h\colon J^1\pi^*\to{\cal M}\pi$ of the projection
 $\mu$ is called a {\sl Hamiltonian section}.
 The differentiable forms
 $\Theta_{h}:=h^*\Theta$ and $\Omega_{h}:=-{\rm d}\Theta_{h}=h^*\Omega$
 are called the {\rm Hamilton-Cartan $k$ and $(k+1)$ forms} of $J^1\pi^*$
 associated with the Hamiltonian section $h$.
$(J^1\pi^*,h)$ is said to be a {\sl Hamiltonian system} in $J^1\pi^*$.
\end{definition}

 In natural coordinates we have that
 $h(t^A,q^i,p^A_i)= (t^A,q^i,p^A_i,p=-{\cal H}(t^A,q^i,p^A_i))$,
and ${\cal H}\in C^\infty  (U)$, $U\subset J^1\pi^*$,
is a {\sl local Hamiltonian function}. Then we have
$$
 \Theta_h = p_i^A{\rm d} q^i\wedge{\rm d}^{k-1}t_A -{\cal H}{\rm
 d}^kt
 \ , \
 \Omega_h = -{\rm d} p_i^A\wedge{\rm d} q^i\wedge{\rm d}^{k-1}t_A +
 {\rm d}{\cal H}\wedge{\rm d}^kt \ .
$$

The field equations for these multisymplectic Hamiltonian systems can be stated as
 \begin{equation}
\label{hamsect}
 \psi^*\inn (X)\Omega_h= 0 \quad ,\quad \mbox{\rm for
every  $X\in\vf(J^1\pi^*)$} \ ,
 \end{equation}
 where $\bar\psi\colon M\to J^1\pi^*$ are sections of the projection
 $\bar\sigma$ that are
solutions to these equations. In natural coordinates, writing
$\bar\psi(t)=(t,\bar\psi^i(t),\bar\psi^A_i(t))$, we have that this
equation is equivalent to the {\sl Hamilton-de Donder-Weyl
equations} for the multisymplectic Hamiltonian system $(J^1\pi^*,h)$
 \begin{equation}
 \label{hem}
  \frac{\partial {\cal H}}{\partial q^i}=-\ds\sum_{A=1}^k\frac{\partial\bar\psi^A_i}{\partial t^A}
 \quad , \quad
 \frac{\partial {\cal H}}{\partial p^A_i}=\frac{\partial\bar\psi^i}{\partial t^A}
 \quad ; \quad 1\leq A\leq k, \, 1\leq\ i \leq n \, .
\end{equation}

We denote by $\vf^k_h(J^1\pi^*)$ the set of
$k$-vector fields ${\bf \bar X}=(\bar X_1,\ldots,\bar X_k)$ in $J^1\pi^*$
 which are solution to the equations
 \begin{equation}
 \inn ({\bf\bar X})\Omega_h=\inn(\bar X_1)\ldots\inn(\bar X_k)\Omega_h=0
 \quad , \quad
\inn({\bf\bar X})\omega=
 \inn(\bar X_1)\ldots\inn(\bar X_k)\omega=1 \ ,
 \label{hameq1}
\end{equation}
(we denote by $\omega=\d^kt$ the volume form in $M$ and
its pull-backs to all the manifolds.
The contraction of $k$-vector fields and forms is the usual one
between tensorial objects).

In a system of natural coordinates, the components of  ${\bf\bar X}$
are given by (\ref{xcoor}),
then $\inn ({\bf\bar X})\omega=1$ leads to $(\bar X_A)^B=1$,
for every $A,B=1,\dots,k$,
and hence the other equation (\ref{hameq1}) gives
 \begin{equation}
\derpar{{\cal H}}{q^i}=-\ds\sum_{A=1}^k (\bar X_A)^A_i
 \quad , \quad
 \derpar{{\cal H}}{p_i^A}= (\bar X_A)^i \  .
 \label{eqsG2}
 \end{equation}

The existence of $k$-vector fields that are solutions to (\ref{hameq1})
 is assured, and in a local system of coordinates they depend on
$n(k^2-1)$ arbitrary functions, but the number of arbitrary functions
for integrable solutions is, in general, less than $n(k^2-1)$.

\begin{proposition}
Let ${\bf\bar X}=(\bar X_1,\dots,\bar X_k)$ be an integrable $k$-vector field in $J^1\pi^*$
and $\bar\psi\colon M\to J^1\pi^*$
an integral section of ${\bf \bar X}$. Then
$\bar \psi(t)=(t,\bar\psi^i(t),\bar\psi^A_i(t))$ is a solution
 to the equations (\ref{hem}), and hence to (\ref{hamsect}), if, and only if,
${\bf \bar X}\in\vf^k_h(J^1\pi^*)$.
\end{proposition}
\proof
If $\bar\psi(t)=(t,\bar\psi^i(t),\bar\psi^A_i(t))$ is an
integral section of ${\bf \bar X}$, we have that
 \begin{equation}
\frac{\partial \bar\psi^i}{\partial t^B}=(\bar X_B)^i \quad ,
\quad \frac{\partial \bar\psi^A_i}{\partial t^B}=(\bar X_B)^A_i \, ,
\label{intsec00}
 \end{equation}
and therefore we obtain that (\ref{eqsG2}) are the HDW-equations
(\ref{hem}). \qed

\subsection{Relation with the $k$-cosymplectic Hamiltonian formalism}
\protect\label{rkcf}

In order to compare the multisymplectic and the $k$-cosymplectic formalisms
of field theory, from now on we consider the case
when $\pi\colon E\to M$ is the trivial bundle
$\rk\times Q \rightarrow \rk$.
Then we can establish relations among the canonical
multisymplectic form on ${\cal M}\pi\equiv\Lambda_2^kT^*(\rk\times Q)$,
the canonical $k$-symplectic structure on $(T^1_k)^*Q$,
and the canonical $k$-cosymplectic structure on $\rk\times(T^1_k)^*Q$
(see also \cite{LMNRS-2002}).
First recall that in $M=\rk$ we have the canonical volume form
 $\omega=\d t^1\wedge\ldots\wedge\d t^k\equiv{\rm d}^kt$.
Then:

\begin{proposition}
\ben
\item
${\cal M}\pi\equiv\Lambda_2^kT^*(\rk\times Q)$
 is diffeomorphic to $\rk\times\Real\times (T^1_k)^*Q$.
\item
$J^1\pi^*$ is diffeomorphic to $\rk\times (T^1_k)^*Q$.
\een
\end{proposition}
\proof

  \begin{enumerate}
\item
Consider the canonical embedding $\imath_t\colon Q\hookrightarrow
\rk\times Q$ given by $i_t(q)=(t,q)$, and the canonical submersion
$\rho_2\colon\rk\times Q\rightarrow Q$.
 We can define the map
$$
\begin{array}{cccc}
\bar\Psi\colon& \Lambda_2^kT^*(\rk\times Q) & \longrightarrow & \rk\times\Real\times (T^1_k)^*Q \\
 & \alpha_{(t,q)} & \mapsto & (t,p,\alpha^1_q, \dots ,\alpha^k_q)
\end{array}
$$
where 
 \begin{eqnarray*}
p&=&\alpha_{(t,q)}\left(\derpar{}{t^1}\Big\vert_{(t,q)},\dots,\derpar{}{t^k}
\Big\vert_{(t,q)}\right)
\\
\alpha^A_q(X) &=&
\alpha_{(t,q)}\left(\derpar{}{t^1}\Big\vert_{(t,q)},\dots,
\derpar{}{t^{A-1}}\Big\vert_{(t,q)},(\imath_t)_*X,
 \derpar{}{t^{A+1}}\Big\vert_{(t,q)},\ldots,
\derpar{}{t^k}\Big\vert_{(t,q)}\right)
 \ ,  \ X\in\vf(Q)
      \end{eqnarray*}
(note that $t^A$ and $p$ are now global coordinates in the
corresponding fibres). The inverse of $\bar\Psi$ is given by
$$
\alpha_{(t,q)}= p \, \d^k
t\vert_{(t,x)}+(\rho_2)_{(t,q)}^*\alpha^A_q\wedge\d^{k-1}t_A\vert_{(t,q)}
\ .
$$
Thus, $\bar\Psi$ is a diffeomorphism. Locally $\bar\Psi$ is written
as the identity.
\item
It is a straighforward consequence of the above item because
$$
J^1\pi^*=\Lambda_2^kT^*E/\pi^*\Lambda^kT^*M\simeq
\rk\times\Real\times (T^1_k)^*Q/\Real\simeq\rk\times (T^1_k)^*Q
$$
\end{enumerate}
\qed

Next, using a procedure analogous to that in the above proof,
 we can give the

\bigskip

 {\bf Relationship between the canonical geometric structures in
$\rk\times\Real\times (T^1_k)^*Q$ and in $(T^1_k)^*Q$}.

\bigskip

 Let $\jmath\colon (T^1_k)^*Q\hookrightarrow\rk\times\Real\times (T^1_k)^*Q$
 be the natural embedding of $(T^1_k)^*Q$ into $\rk\times\Real\times (T^1_k)^*Q$
as the zero-section of the bundle $\rk\times\Real\times (T^1_k)^*Q\to (T^1_k)^*Q$.
Starting from the canonical forms $\Theta$ and $\Omega$ in
${\cal M}\pi\simeq\rk\times\Real\times (T^1_k)^*Q$
we can define the forms $\theta^A$ on $(T^1_k)^*Q$, $1\leq A \leq k$, by
 \beann
 \theta^A(X)&=&\jmath^*\left[\Theta\left(\derpar{}{t^1},\dots,
\derpar{}{t^{A-1}},\jmath_*X,\derpar{}{t^{A+1}},\ldots,\derpar{}{t^k}\right)\right]
\\ &=&
 -\left(\jmath^*\left[\inn\Big(\ds\derpar{}{t^k}\Big)\ldots
 \inn\Big(\ds\derpar{}{t^1}\Big)(\Theta\wedge\d t^A)\right]\right)(X)
 \quad , \quad X\in\vf((T^1_k)^*Q)\ .
 \eeann
 Then for $X,Y\in\vf( (T^1_k)^*Q)$,
 we get the $2$-forms $\omega^A$ on $(T^1_k)^*Q$ given as
 \bea
\omega^A(X,Y)&=&-\d\theta^A(X,Y)=
\jmath^*\left[\Omega\left(\jmath_*X,\derpar{}{t^1},\dots,
\derpar{}{t^{A-1}},\jmath_*Y,\derpar{}{t^{A+1}},\ldots,\derpar{}{t^k}\right)\right]
\nonumber \\ &=&
 (-1)^{k+1}\left(\jmath^*\left[\inn\Big(\ds\derpar{}{t^k}\Big)\ldots
 \inn\Big(\ds\derpar{}{t^1}\Big)(\Omega\wedge\d
 t^A)\right]\right)(X,Y)\, .
 \label{relatomega1}
\eea
  From (\ref{coormult}) we obtain the local expressions
 $$
\theta^A =p^A_i \d q^i \quad , \quad
\omega^A = \d q^i \wedge\d p^A_i  \ .
$$
Furthermore, we have the involutive distribution
$\ds V=\ker\,(\pi^1_Q)_*$,
 and hence $(\omega^A,V;1\leq A \leq k)$
is the canonical $k$-symplectic structure in $(T^1_k)^*Q$.

Conversely, starting from this $k$-symplectic structure in $(T^1_k)^*Q$
we can obtain the canonical forms in ${\cal M}\pi\simeq\rk\times\Real\times (T^1_k)^*Q$,
by doing
\begin{equation}
\Theta= p \d^k t+\sigma_2^*\theta^A\wedge\d^{k-1}t_A \quad , \quad
\Omega=-\d\Theta=-\d p\wedge\d^k t+\sigma_2^*\omega^A\wedge\d^{k-1}t_A
 \label{relatomega2}
\end{equation}
where $\sigma_2\colon\rk\times\Real\times (T^1_k)^*Q\to (T^1_k)^*Q$
is the canonical submersion.

Summarizing, we have proved that:

\begin{theorem}
The canonical multisymplectic form on
${\cal M}\pi\simeq\rk\times\Real\times (T^1_k)^*Q$ and
the $2$-forms of the canonical $k$-symplectic structure on $(T^1_k)^*Q$
are related by (\ref{relatomega1}), and (\ref{relatomega2}).
\end{theorem}

\bigskip

 {\bf Relationship between the canonical geometric structures in
$\rk\times\Real\times (T^1_k)^*Q$ and in $\rk \times(T^1_k)^*Q$}.

\bigskip

In an analogous way, we can also relate the canonical geometric
structures in $\rk\times\Real\times (T^1_k)^*Q$ and in $\rk\times
(T^1_k)^*Q$. In fact, denoting by $\mathfrak{i}\colon
\rk\times(T^1_k)^*Q\hookrightarrow\rk\times\Real\times (T^1_k)^*Q$
 the natural embedding of $\rk\times(T^1_k)^*Q$ into $\rk\times\Real\times (T^1_k)^*Q$
as the zero-section of the bundle $\mu\colon\rk\times\Real\times (T^1_k)^*Q\to\rk\times (T^1_k)^*Q$;
then from the canonical forms $\Theta$ and $\Omega$ in
${\cal M}\pi\simeq\rk\times\Real\times (T^1_k)^*Q$
we can define the forms $\Theta^A$ on $\rk\times(T^1_k)^*Q$
 as follows:
for $\bar X\in\vf(\rk\times(T^1_k)^*Q)$, and $1\leq A \leq k$,
 \beann
 \Theta^A(\bar X)&=&\mathfrak{i}^*\left[\Theta\left(\derpar{}{t^1},\dots,
\derpar{}{t^{A-1}},\mathfrak{i}_*\bar
X,\derpar{}{t^{A+1}},\ldots,\derpar{}{t^k}\right)\right]
\\ &=&
-\left(\mathfrak{i}^*\left[\inn\Big(\ds\derpar{}{t^k}\Big)\ldots
 \inn\Big(\ds\derpar{}{t^1}\Big)(\Theta\wedge\d t^A)\right]\right)(\bar X)
 \eeann
 Then, for $\bar X,\bar Y\in\vf(\rk\times(T^1_k)^*Q)$,
 we obtain the $2$-forms $\Omega^A$ on $\rk\times(T^1_k)^*Q$,
 \bea
 \Omega^A(\bar X,\bar Y)&=&-\d\Theta^A(\bar X,\bar Y)=
\mathfrak{i}^*\left[\Omega\left(\mathfrak{i}_*\bar X,\derpar{}{t^1},\dots,
\derpar{}{t^{A-1}},\mathfrak{i}_*\bar
Y,\derpar{}{t^{A+1}},\ldots,\derpar{}{t^k}\right)\right]
 \nonumber
\\ &=& (-1)^{k+1}\left(\mathfrak{i}^*\left[i\Big(\ds\derpar{}{t^k}\Big)\ldots
 i\Big(\ds\derpar{}{t^1}\Big)(\Omega\wedge\d t^A)\right]\right)(\bar X,\bar Y)
\, .
 \label{relatomega3}
\eea
(These forms have the same coordinate expressions as $\theta^A$ and $\omega^A$).
Furthermore, although the $1$-forms $\eta^A$
are canonically defined on $\rk\times(T^1_k)^*Q$, we can recover them
from the multisymplectic form $\Omega$ as follows:
for $\bar X\in\vf(\rk\times(T^1_k)^*Q)$,
 \begin{equation}
 \eta^A(\bar X)=(-1)^{k-A}\mathfrak{i}^*\left[\Omega\left(\derpar{}{p},\derpar{}{t^1},\dots,
\derpar{}{t^{A-1}},\mathfrak{i}_*\bar X,\derpar{}{t^{A+1}},\ldots,
\derpar{}{t^k}\right)\right]  \ .
 \label{etas}
\end{equation}
whose coordinate expressions are $\eta^A=\d t^A$.
These forms can also be defined by introducing the canonical embedding
$$
\begin{array}{cccc}
\jmath_0\colon & \rk\times (T^1_k)^*Q & \hookrightarrow &
 \rk\times\Real\times (T^1_k)^*Q \\
 & (t,\alpha^1_x, \dots ,\alpha^k_x) & \rightarrow & (t,1,0_x, \dots ,0_x)
\end{array}
$$
and then making
\begin{equation}
 \eta^A(\bar X)=\jmath_0^*\left[\Theta\left(\derpar{}{t^1},\dots,
\derpar{}{t^{A-1}},(\jmath_0)_*\bar X,\derpar{}{t^{A+1}},\ldots,
\derpar{}{t^k}\right)\right]  \,,   \quad  \bar
X\in\vf(\rk\times(T^1_k)^*Q)\ .
 \label{etas2}
\end{equation}
Furthermore, we have the involutive distribution
$\ds{\cal V}=\ker\,(\bar\pi_2)_*=\left\langle\derpar{}{t^A}\right\rangle$,
and hence $(\eta^A,\Omega^A,{\cal V};1\leq A \leq k)$
is the canonical $k$-cosymplectic structure in $\rk\times(T^1_k)^*Q$.

Conversely, starting from this $k$-cosymplectic structure in $\rk\times(T^1_k)^*Q$
we can obtain the canonical forms in ${\cal M}\pi\simeq\rk\times\Real\times (T^1_k)^*Q$,
by doing
\begin{equation}
\Theta= p \d^k t+\bar\sigma_2^*\Theta^A\wedge\d^{k-1}t_A \quad , \quad
\Omega=-\d\Theta=-\d p\wedge\d^k t+\bar\sigma_2^*\Omega^A\wedge\d^{k-1}t_A
 \label{relatomega4}
\end{equation}
where $\bar\sigma_2\colon\rk\times\Real\times (T^1_k)^*Q\to (T^1_k)^*Q$
is the canonical submersion.

Summarizing, we have proved that:

\begin{theorem}
The canonical multisymplectic form on ${\cal
M}\pi\simeq\rk\times\Real\times (T^1_k)^*Q$ and the $1$ and
$2$-forms of the canonical $k$-cosymplectic structure on $\rk\times
(T^1_k)^*Q$ are related by (\ref{relatomega3}), (\ref{etas}) (or
(\ref{etas2})), and (\ref{relatomega4}).
\end{theorem}

\bigskip

 {\bf Relationship between the canonical geometric structures in
$J^1\pi^*\simeq\rk\times  (T^1_k)^*Q$.}

\bigskip

It is important to point out that, as the bundle
$\mu\colon{\cal M}\pi\simeq\rk\times\Real\times (T^1_k)^*Q\to
 J^1\pi^*\simeq\rk\times (T^1_k)^*Q$
is trivial, then Hamiltonian sections can be taken to be global
sections of the projection $\mu$ by giving a global Hamiltonian
function ${\rm H}\in C^\infty(\rk\times (T^1_k)^*Q)$. Then we can
also relate the non-canonical multisymplectic form with the
$k$-cosymplectic structure in $\rk\times (T^1_k)^*Q$ as follows:
starting from the forms $\Theta_h$ and $\Omega_h$ in
$\rk\times(T^1_k)^*Q$, we can define the forms $\Theta^A$ and
$\Omega^A$ on $\rk\times(T^1_k)^*Q$ as follows:
for $\bar X,\bar Y\in\vf(\rk\times(T^1_k)^*Q)$, and $1\leq A \leq k$,
 \bea
 \Theta^A(\bar X) &=& -
\left(\inn\left(\derpar{}{t^k}\right)\ldots
\inn\left(\derpar{}{t^1}\right)(\Theta_h\wedge\d t^A)\right)(\bar X)
\nonumber \\
\Omega^A(\bar X,\bar Y)&=&-\d\Theta^A(\bar X,\bar Y)=
 (-1)^{k+1}\left(\inn\left(\derpar{}{t^k}\right)\ldots
\inn\left(\derpar{}{t^1}\right)(\Omega_h\wedge\d t^A)\right)(\bar X,\bar Y) \ ,
 \label{relatomega5}
\eea
and the $1$-forms $\eta^A=\d t^A$ are canonically defined.

Conversely, starting from the canonical $k$-cosymplectic structure on $\rk\times(T^1_k)^*Q$,
and from ${\cal H}$, we can construct
\begin{equation}
\Theta_h= -{\cal H} \d^k t+\Theta^A\wedge\d^{k-1}t_A \quad , \quad
\Omega=-\d\Theta=\d{\cal H}\wedge\d^k t+\Omega^A\wedge\d^{k-1}t_A
 \label{relatomega6}
\end{equation}

So we have:

\begin{theorem}
The multisymplectic form and the $2$-forms of the canonical
$k$-cosymplectic structure  on $J^1\pi^*\simeq\rk\times (T^1_k)^*Q$
are related by (\ref{relatomega5}) and (\ref{relatomega6}).
\end{theorem}

Finally, the following result about the solutions to the Hamiltonian equations
establishes the complete equivalence between both formalisms:

\begin{theorem}
A $k$-vector field ${\bf \bar X}=(\bar X_1,\ldots,\bar X_k)$
 in $J^1\pi^*\simeq\rk\times (T^1_k)^*Q$ is a solution to the equations
(\ref{geonah}) if, and only if, it is also a solution to the equations (\ref{hameq1});
that is, $\vf^k_h(\rk\times (T^1_k)^*Q)=\vf^k_H(\rk\times (T^1_k)^*Q)$.
\end{theorem}
\proof
The proof is immediate, bearing in mind that
in natural coordinates the solutions to the equations (\ref{geonah}) and (\ref{hameq1})
are partially determined by the equations (\ref{111}) and (\ref{eqsG2}) respectively,
and these are equivalent.
\qed

\section{Multisymplectic Lagrangian formalism}

\subsection{Multisymplectic Lagrangian systems}
\protect\label{mls}

(For details, see \cite{EMR-98} and the references quoted therein).
Consider the first-order jet bundle $\pi_E\colon J^1\pi\to E$, which is also a
bundle over $M$ with projection $\bar\pi\colon J^1\pi\longrightarrow
M$, and is endowed with natural coordinates $(t^A,q^i,v^i_A)$,
adapted to the bundle structure. A {\sl Lagrangian density} is a
$\bar\pi$-semibasic $k$-form on $J^1\pi$, and hence it can be
expressed as ${\Lagd} =\Lag\omega$, where ${\Lag}\in
C^{\infty}(J^1\pi)$ is the {\sl Lagrangian function} associated with
$\Lagd$ and $\omega$. Using the canonical structures of $J^1\pi$, we
can define the {\sl Poincar\'e-Cartan $k$ and $(k+1)$-forms},
which have the following local expressions:
\beann
\Theta_{\Lagd}&=&\derpar{\Lag}{v^i_A}\d q^i\wedge\d^{k-1}t_A -
\left(\derpar{\Lag}{v^i_A}v^i_A -\Lag\right)\d^kt
\\
\Omega_{\Lagd}&=&
-\frac{\partial^2\Lag}{\partial v^j_B\partial v^i_A}
\d v^j_B\wedge\d q^i\wedge\d^{k-1}t_A
-\frac{\partial^2\Lag}{\partial q^j\partial v^i_A}\d q^j\wedge\d q^i\wedge\d^{k-1}t_A +
 \\  & &
\frac{\partial^2\Lag}{\partial v^j_B\partial v^i_A}v^i_A\d v^j_B\wedge\d^kt  +
\left(\frac{\partial^2\Lag}{\partial q^i\partial v^j_B}v^j_B
 -\derpar{\Lag}{q^j}+\frac{\partial^2\Lag}{\partial t^A\partial v^i_A}\right)\d q^i\wedge\d^kt
\eeann
$(J^1\pi,\Lagd)$ is said to be a {\rm Lagrangian system}.
The Lagrangian system and the Lagrangian function are
{\sl regular} if $\Omega_{\Lagd}$ is a multisymplectic
$(k+1)$-form. Elsewhere they are {\sl singular} (or {\sl non-regular}),
and $\Omega_{\Lagd}$ is a pre-multisymplectic form.
The regularity condition is locally equivalent to
$det (\frac{\partial^2\Lag}{\partial v^A_\alpha\partial v^B_\nu})\not= 0$,
at every point in $J^1\pi$.

The Lagrangian field equations can be stated as
 $$
(\phi^{1})^*\inn (X)\Omega_{\Lagd}= 0 \quad ,\quad \mbox{\rm for
every $ X\in\vf (J^1\pi)$} \ ,
$$
where $\phi\colon M\to E$ are sections of the projection $\pi$, and
$\phi^{1}\colon M\to J^1\pi$ are their canonical liftings, which are
solutions to these equations. In natural coordinates, writing
$\phi(t)=(t,\phi^i(t))$, we have that this equation is equivalent to
the {\sl Euler-Lagrange equations} (\ref{ELe}) for the Lagrangian
$\Lag$. Furthermore, we denote by $\vf^k_{\Lagd} (J^1\pi)$ the set
of $k$-vector fields ${\bf
\bar\Gamma}=(\bar\Gamma_1,\dots,\bar\Gamma_k)$ in $J^1\pi$,
 that are solutions to the equations
 \begin{equation}
 \inn ({\bf \bar\Gamma})\Omega_{\Lagd}=0 \quad , \quad
 \inn ({\bf \bar\Gamma})\omega=1
 \label{lageq1}
 \end{equation}
In a system of natural coordinates the components of  ${\bf \bar\Gamma}$
are given by (\ref{xlagcoor}),
 then ${\bf \bar\Gamma}$ is a solution to (\ref{lageq1}) if, and only if,
$(\bar\Gamma_A)^B=1$, for every $A,B=1,\ldots,k$, and
$(\bar\Gamma_A)^i$ and $(\bar\Gamma _A)^i_B$ satisfy the equations (\ref{elcoor}).
When $\Lag$ is regular, we obtain that $(\bar\Gamma_A)^i=v^i_A$, and the
equations (\ref{elcoor1} hold;
then ${\bf \bar\Gamma}$ is a {\sc sopde}, and hence,
if it is integrable, its integral sections are holonomic and
they are solutions to the Euler-Lagrange equations for $\Lag$.
If $\Lag $ is not regular, the existence of solutions to the equations (\ref{ELe})
for $\Lag$ or to (\ref{lageq1}) is not assured, in general,
except in a submanifold of $J^1\pi$
(in the most favourable situations). Moreover,
solutions to (\ref{lageq1}) are not {\sc sopde} necessarily.

Finally, $\Theta_{\Lagd}\in\df^1(J^1\pi)$ being $\pi_E$-semibasic,
we have a natural map
$\widetilde{F\Lag}\colon J^1\pi\to{\cal M}\pi$,
given by
   $$
  \widetilde{F\Lag}({\bar y})=\Theta_{\Lagd}({\bar y}) \quad ; \quad \bar y\in J^1\pi
  $$
which is called the {\sl extended Legendre map} associated to the
Lagrangian $\Lag$. The {\sl restricted Legendre map} is
$F{\Lag} =\mu\circ\widetilde{F{\Lag}}\colon J^1\pi\to J^1\pi^*$. Their local expressions are
\bea
\widetilde{F{\Lag}} &\colon& (t^A, q^i,v^i_A)\mapsto
\left( t^A, q^i,\derpar{{\Lag}}{v^i_A},{\Lag}-v^i_A\derpar {{\Lag}}{v^i_A}\right)
 \nonumber \\
F{\Lag} &\colon& (t^A, q^i,v^i_A)\mapsto \left( t^A, q^i,\derpar{{\Lag}}{v^i_A}\right)
\label{legmulti}
\eea
Moreover, we have $\widetilde{F{\Lag}}^*\Theta=\Theta_{\Lagd}$,
and $\widetilde{F{\Lag}}^*\Omega=\Omega_{\Lagd}$.
Observe that the Legendre transformations $F{\Lag}$
defined for the $k$-cosymplectic and the multisymplectic formalisms
are the same, as their local expressions (\ref{locfl1}) and (\ref{legmulti}) show.

\subsection{Relation between multisymplectic and $k$-cosymplectic Lagrangian systems}
\protect\label{rmk}

In the particular case $E=\rk\times Q$, we have $J^1\pi\simeq\rk\times T^1_kQ$
and we can define the Energy Lagrangian function ${\cal E}_{\Lag}$ as
$$
{\cal
E}_{\Lag}=\Theta_{\Lagd}\left(\derpar{}{t^1},\dots,\derpar{}{t^k}\right)
$$
whose local expression is $\ds {\cal E}_{\Lag} =v^i_A\frac{\partial
\Lag}{\partial v^i_A}-\Lag$. Then we can write
$$
\Theta_{\Lagd}=\derpar{\Lag}{v^i_A}\d q^i\wedge\d^{k-1}t_A - {\cal
E}_{\Lag}\d^kt \quad , \quad \Omega_{\Lagd}=
-\d\left(\derpar{\Lag}{v^i_A}\right)\wedge\d q^i\wedge\d^{k-1}t_A +
\d{\cal E}_{\Lag}\wedge\d^kt
$$

In this particular case, as in the Hamiltonian case, we can relate
the non-canonical Lagrangian multisymplectic (or
pre-multisymplectic) form $\Omega_{\Lagd}$ with the non-canonical
Lagrangian $k$-cosymplectic (or $k$-precosymplectic) structure in
$\rk\times T^1_kQ$ constructed in Section \ref{kcolag} as follows:
starting from the forms $\Theta_{\Lagd}$ and $\Omega_{\Lagd}$ in
$J^1\pi \simeq\rk\times T^1_kQ$, we can define the forms
$\Theta_{\Lag}^A$ and $\Omega_{\Lag}^A$ on $\rk\times T^1_kQ$,
as follows:  for $X,Y\in\vf(\rk\times T^1_kQ)$, and $1\leq A \leq k$,
\bea
 \Theta_{\Lag}^A(X) &=& -\left(\inn\left(\derpar{}{t^k}\right)\ldots
\inn\left(\derpar{}{t^1}\right)(\Theta_{\Lagd}\wedge\d t^A)\right)(X)
\nonumber \\
\Omega_{\Lag}^A(X,Y)&=&
-\d\Theta_{\Lag}^A=(-1)^{k+1}\left(\inn\left(\derpar{}{t^k}\right)\ldots
 \inn\left(\derpar{}{t^1}\right)(\Omega_{\Lagd}\wedge\d t^A)\right)(X,Y) \ .
 \label{relatomegal}
\eea
and the $1$-forms $\eta^A=\d t^A$ are canonically defined.

Conversely, starting from the Lagrangian $k$-cosymplectic (or
$k$-precosymplectic) structure on $\rk\times T^1_kQ$, and from
${\cal E}_{\Lag}$, we can construct on $ \rk\times T^1_kQ\simeq
J^1\pi$
\begin{equation}
\Theta_{\Lagd}= -{\cal E}_{\Lag}\d^k t+\Theta^A_{\Lag}\wedge\d^{k-1}t_A \quad , \quad
\Omega_{\Lagd}=-\d\Theta_{\Lagd}=\d{\cal E}_{\Lag}\wedge\d^k t+\Omega_{\Lag}^A\wedge\d^{k-1}t_A
 \label{relatomegal2}
\end{equation}

So we have proved that:

\begin{theorem}
The Lagrangian multisymplectic (or pre-multisymplectic) form and the
Lagrangian $2$-forms of the $k$-cosymplectic (or
$k$-precosymplectic) structure on $J^1\pi\equiv\rk\times  T^1_kQ$
are related by (\ref{relatomegal}) and (\ref{relatomegal2}).
\end{theorem}

The discussion in the above section about the Lagrangian equations
proves the following result, which
establishes the complete equivalence between both formalisms:

\begin{theorem}
A $k$-vector field ${\bf \bar\Gamma}=(\bar\Gamma_1,\ldots,\bar\Gamma_k)$
 in $J^1\pi\simeq\rk\times T^1_kQ$ is a solution to the equations
(\ref{lageq1}) if, and only if, it is also a solution to the equations (\ref{genericELe});
that is, we have that
$\vf^k_{\Lag}(\rk\times T^1_kQ)=\vf^k_{\Lag}(\rk\times T^1_kQ)$.
\end{theorem}

\section*{Appendix: Correspondences between the formalisms}
\begin{center}{\bf Hamiltonian formalism}\end{center}
$$\begin{array}{llll}
    &\makebox{\bf$k$-symplectic} & \makebox{\bf$k$-cosymplectic} & \makebox{\bf Multisymplectic} \\
    \noalign{\medskip}
\mbox{Phase space}
& (T^1_k)^*Q & \rk\times  (T^1_k)^*Q&  {\cal M}\pi\to J^1\pi^*\\ \noalign{\medskip}
   \begin{array}{l} \mbox{Canonical} \\ \mbox{forms}\end{array}&
\theta^A\in\Lambda^1((T^1_k)^*Q ) &
\Theta^A\in\Lambda^1(\rk\times(T^1_k)^*Q )
 &  \Theta\in\Lambda^k({\cal M}\pi)\\ \noalign{\medskip}
   &\omega^A=-d\theta^A  & \Omega^A=-d\Theta^A & \Omega=-d\Theta \\ \noalign{\medskip}
 \mbox{Hamiltonians}  &H:(T^1_k)^*Q\to \r  & {\cal H}:\rk\times (T^1_k)^*Q\to \r &
 h:J^1\pi^* \to {\cal M}\pi \\ \noalign{\medskip}
   &  &  & \Theta_h=h^*\Theta \, , \,  \Omega_h=h^*\Omega\\
\noalign{\medskip}
  \begin{array}{l} \mbox{Geometric}\\ \mbox{equations}\end{array}  &
 {\ds\sum_{A=1}^ki(X_A)\omega^A=dH} &
   \begin{array}{c}    \ds\sum_{A=1}^ki(\bar{X}_A)
  \Omega^A=dH-\frac{\partial
  H}{\partial  t^A}dt^A   \\ dt^A(\bar{X}_B)=\delta^A_B \end{array} &
 \begin{array}{c}  \inn ({\bf \bar X})\Omega_h=0 \\
 \inn ({\bf \bar X})\omega=1 \end{array}
\\ \noalign{\medskip}
   &\begin{array}{c}(X_1,\ldots,X_k)  \\ \mbox{k-vector field on} \,
   (T^1_k)^*Q\end{array}&\begin{array}{c}(\bar{X}_1,\ldots,\bar{X}_k)
    \\ \mbox{k-vector field on} \, \rk\times (T^1_k)^*Q\end{array}  &
{\bf\bar X} \ \mbox{k-vector field on}\ J^1\pi^*
  \end{array}$$

\bigskip

\begin{center}{\bf Lagrangian  formalism}\end{center}
$$\begin{array}{llll}
    &\makebox{\bf$k$-symplectic} & \makebox{\bf$k$-cosymplectic} & \makebox{\bf Multisymplectic} \\
    \noalign{\medskip}
  \mbox{Phase space} & T^1_kQ & \rk\times  T^1_kQ&  J^1\pi\\ \noalign{\medskip}
  \mbox{Lagrangians}  &L:T^1_kQ\to \r  & {\cal L}:\rk\times T^1_kQ\to \r &
 {\cal L}:J^1\pi \to \r \, , \, {\Lagd}={\cal L}\omega\\ \noalign{\medskip}
\begin{array}{l} \mbox{Lagrangian} \\ \mbox{forms}\end{array}& \theta^A_L\in\Lambda^1(T^1_kQ ) &
\Theta^A_{{\cal L}}\in\Lambda^1(\rk\times T^1_kQ )
 &  \Theta_{\Lagd}\in\Lambda^k(J^1\pi)\\ \noalign{\medskip}
   &\omega^A_L=-d\theta^A  & \Omega^A_{{\cal L}}=-d\Theta^A_{{\cal L}} & \Omega_{\Lagd}
   =-d\Theta_{\Lagd} \\ \noalign{\medskip}
  \begin{array}{l} \mbox{Geometric}\\ \mbox{equations}\end{array} &   \ds\sum_{A=1}^ki(\Gamma_A)\omega^A_L=E_L  &
   \begin{array}{c} \ds\sum_{A=1}^ki(\bar\Gamma_A)
  \Omega^A_{\cal L}=d{{\cal E_L}}-\frac{\partial
  {\cal L}}{\partial  t^A}dt^A \\ dt^A(\bar\Gamma_B)=\delta^A_B\end{array} &
 \begin{array}{c}  \inn ({\bf \bar\Gamma})\Omega_{\Lagd}=0 \\
 \inn ({\bf \bar\Gamma})\omega=1 \end{array}
 \\ \noalign{\medskip}
   &\begin{array}{c}(\Gamma_1,\ldots,\Gamma_k) \\ \mbox{k-vector field on}
   \,
   T^1_kQ\end{array}&\begin{array}{c}(\bar\Gamma_1,\ldots,\bar\Gamma_k)
    \\ \mbox{k-vector field on} \, \rk\times T^1_kQ\end{array}  &
     {\bf\bar \Gamma} \ \mbox{k-vector field on}\ J^1\pi
  \end{array}$$

\subsection*{Acknowledgments}

We acknowledge the financial support of the project
MTM2006-27467-E/. NRR also acknowledges the financial support of
{\sl Ministerio de Educaci\'on y Ciencia}, Project MTM2005-04947.
We thank Mr. Jeff Palmer for his assistance in preparing the English
version of the manuscript.

\end{document}